\documentclass[a4paper,english,reprint,superscriptaddress,nofootinbib]{revtex4-1}
\usepackage[T1]{fontenc}
\usepackage[latin9]{inputenc}
\usepackage{verbatim}
\usepackage{amsmath}
\usepackage{amssymb}
\usepackage{cancel}
\usepackage{graphicx}

\makeatletter

\pdfpageheight\paperheight
\pdfpagewidth\paperwidth

\usepackage{babel}

\usepackage[normalem]{ulem}
\newcommand{\FRcrossOut}[1]{}  

\usepackage{babel}

\makeatother

\usepackage{babel}
\begin{document}
\title{Can endogenous fluctuations persist in high-diversity ecosystems?}
\author{Felix Roy}
\affiliation{Institut de physique théorique, Université Paris Saclay, CEA, CNRS,
F-91191 Gif-sur-Yvette, France}
\affiliation{Laboratoire de Physique de l'Ecole Normale Superieure, ENS, Université
PSL, CNRS, Sorbonne Université, Université Paris-Diderot, Sorbonne
Paris Cité, Paris, France}
\author{Matthieu Barbier}
\affiliation{Centre for Biodiversity Theory and Modelling, Theoretical and Experimental
Ecology Station, CNRS and Paul Sabatier University, 09200 Moulis,
France}
\author{Giulio Biroli}
\affiliation{Laboratoire de Physique de l'Ecole Normale Superieure, ENS, Université
PSL, CNRS, Sorbonne Université, Université Paris-Diderot, Sorbonne
Paris Cité, Paris, France}
\author{Guy Bunin}
\affiliation{Department of Physics, Technion-Israel Institute of Technology, Haifa
32000, Israel}
\date{\today}
\begin{abstract}
When can complex ecological interactions drive an entire ecosystem
into a persistent non-equilibrium state, where species abundances
keep fluctuating without going to extinction? We show that high-diversity
spatially-extended systems, in which conditions vary somewhat between
spatial locations, can exhibit chaotic dynamics which persist for
extremely long times. We develop a theoretical framework, based on
dynamical mean-field theory, to quantify the conditions under which
these fluctuating states exist, and predict their properties. We uncover
parallels with the persistence of externally-perturbed ecosystems,
such as the role of perturbation strength, synchrony and correlation
time. But uniquely to endogenous fluctuations, these properties arise
from the species dynamics themselves, creating feedback loops between
perturbation and response. A key result is that the fluctuation amplitude
and species diversity are tightly linked, in particular fluctuations
enable dramatically more species to coexist than at equilibrium in
the very same system. Our findings highlight crucial differences between
well-mixed and spatially-extended systems, with implications for experiments
and their ability to reproduce natural dynamics. They shed light on
the maintenance of biodiversity, and the strength and synchrony of
fluctuations observed in natural systems.
\end{abstract}
\maketitle
\global\long\def\mean{\operatorname{mean}}%
\global\long\def\var{\operatorname{var}}%
\global\long\def\std{\operatorname{std}}%
\global\long\def\corr{\operatorname{corr}}%
\global\long\def\cov{\operatorname{cov}}%

While large temporal variations are widespread in natural populations
\cite{lundberg_population_2000,inchausti_relation_2003}, it is difficult
to ascertain how much they are caused by external perturbations, or
by the ecosystem's internal dynamics, see e.g. \citep{ellner_chaos_1995,scheffer_why_2003}.
In particular, both theoretical tools and empirical results come short
of addressing a fundamental question: can we identify when fluctuations
in species abundances arise from complex ecological interactions? 

Our focus here is on high-diversity communities. Historically, studies
of endogenous fluctuations have focused on single populations or few
species \cite{may_biological_1974,allen_chaos_1993}. On the other
hand, theories of many-species interaction networks often center on
ecosystems that return to equilibrium in the absence of perturbations
\citep{may_will_1972}. Some authors have even proposed that endogenous
fluctuations are generally too rare or short-lived to matter, since
they can be self-defeating: dynamics that lead to large erratic variations
cause extinctions, leaving only species whose interactions are less
destabilizing, with extinctions continuing until an equilibrium is
reached \citep{berryman_are_1989,nisbet_avoiding_1989}.

Many-species endogenous fluctuations can only persist if they do not
induce too many extinctions. Extinction rates are related to the amplitude
of fluctuations \cite{giles_leigh_average_1981,lande_risks_1993},
their synchrony \cite{loreau_biodiversity_2003} and their correlation
time \cite{ripa_noise_1996}. The peculiarity of endogenous fluctuations
is that these properties arise from the species dynamics, and therefore
feed back on themselves. A theory of these feedbacks is however lacking.

Here we propose a novel quantitative approach, and show that many-species
endogenous fluctuations can persist for extremely long times. Furthermore,
they can be realized in experimental conditions, and identified in
these experiments by multiple characteristic features. Crucially,
we show that states with higher species diversity have stronger fluctuations,
and vice versa. We also offer reasons why they may not have been observed
in previous studies, and directions in which to search. An important
factor in maintaining a dynamically fluctuating state is the spatial
extension of the ecosystem, here modeled as a metacommunity: multiple
patches (locations in space) that are coupled by migration.

\begin{figure}
\begin{centering}
\includegraphics[width=0.8\linewidth]{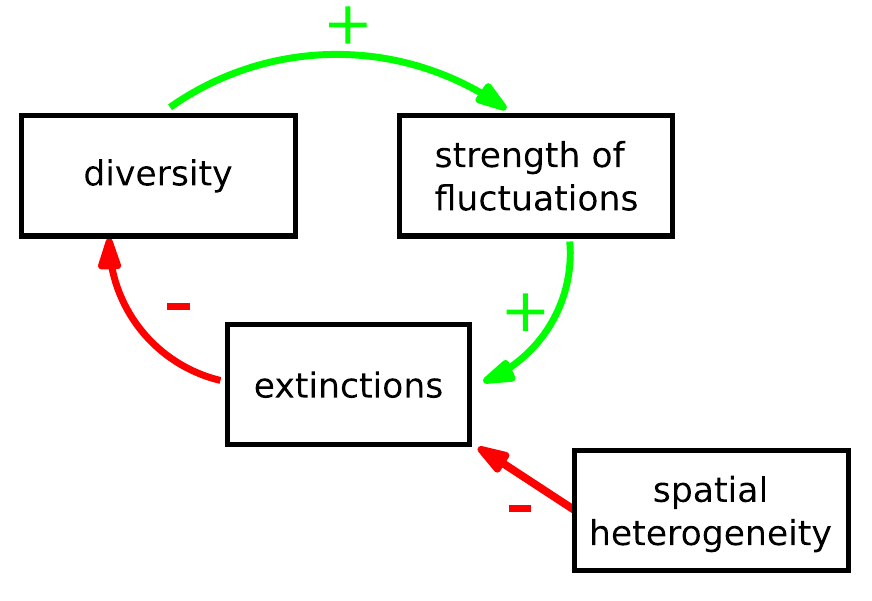} 
\par\end{centering}
\caption{The fluctuation-diversity feedback cycle. Species diversity is required
to maintain endogenous fluctuations. But these fluctuations cause
extinctions, which reduce diversity. This negative feedback cycle
can lead to the disappearance of endogenous fluctuations, especially
in a well-mixed community. However, if spatial heterogeneity can limit
extinctions, this negative feedback loop may slow down and create
a fluctuating state that persists for very long times. \label{fig:feedback_intro}}
\end{figure}

Our strategy is the following. We first propose and simulate experiments
to show that persistent fluctuations can be very elusive in a single
well-mixed community, yet attainable in a metacommunity via three
main ingredients: the existence of multiple patches, moderate migration
fluxes coupling them, and differences in conditions between patches.
These three ingredients can mitigate the likelihood that large fluctuations
within a patch will lead to overall extinctions (see Fig. \ref{fig:feedback_intro}),
and make it possible for species to persist in highly fluctuating
states.%

We then offer a quantitative understanding of this phenomenon. We
build on the analytical framework developed in \cite{roy_numerical_2019}
(dynamical mean-field theory) that allows us to investigate, in a
quantitative and predictive way, the conditions under which robust
fluctuations can arise from complex interactions. This theory exactly
maps a deterministic metacommunity (many-species dynamics over multiple
spatial locations) to a \textit{stochastic representative metapopulation}
(single-species dynamics over multiple spatial locations). It predicts
the distribution of abundance, survival and variability for a species
subjected to ``noise'' that results from other species in the same
community, rather than external perturbations. Dynamical mean-field
theory allows us to analyze these fluctuations, and show that the
effective stochasticity of species dynamics is a manifestation of
high-dimensional chaos.

The intuitive picture that emerges from our analysis is the following:
the persistence of endogenous fluctuations, which can be found in
a wide range of realistic conditions, requires a balancing act between
forces that stabilize and destabilize the dynamics, see Fig. \ref{fig:feedback_intro}.
On the one hand, the system needs to preserve a high diversity (both
in terms of species number and interaction heterogeneity), as it is
known \cite{may_will_1972,gravel_stability_2016} that lower diversity
leads to a stable equilibrium. On the other hand, the system also
has to limit excursions towards very low abundances. This requires
weeding out species that induce unsustainable fluctuations, and rescuing
the others from sudden drops.

To accomplish that, the system relies on asynchronous dynamics between
different spatial locations, and finite strength and correlation time
of the abundance fluctuations. In addition, despite large fluctuations
in the abundances of all species, that strongly affect the species
growth in any given patch, long-lasting ``sources'' emerge for some
of the species, i.e. patches where these species are more likely to
remain away from extinction. Rare dynamical fluctuations leading to
extinction in a given patch are hampered by migration from the other
patches, which keeps the system in a non-equilibrium state. We show
that, with moderate migration and some spatial heterogeneity, high-diversity
dynamical states can be reached where species populations fluctuate
over orders of magnitude, yet remain bounded for very long times above
their extinction threshold.

Our findings allow us to paint a more precise picture of when persistent
endogenous fluctuations can arise. We conclude with a discussion of
the implications for biodiversity and ecosystem stability, and predictions
for future experiments on community dynamics.%

\begin{figure*}
\begin{centering}
\includegraphics[width=0.8\linewidth]{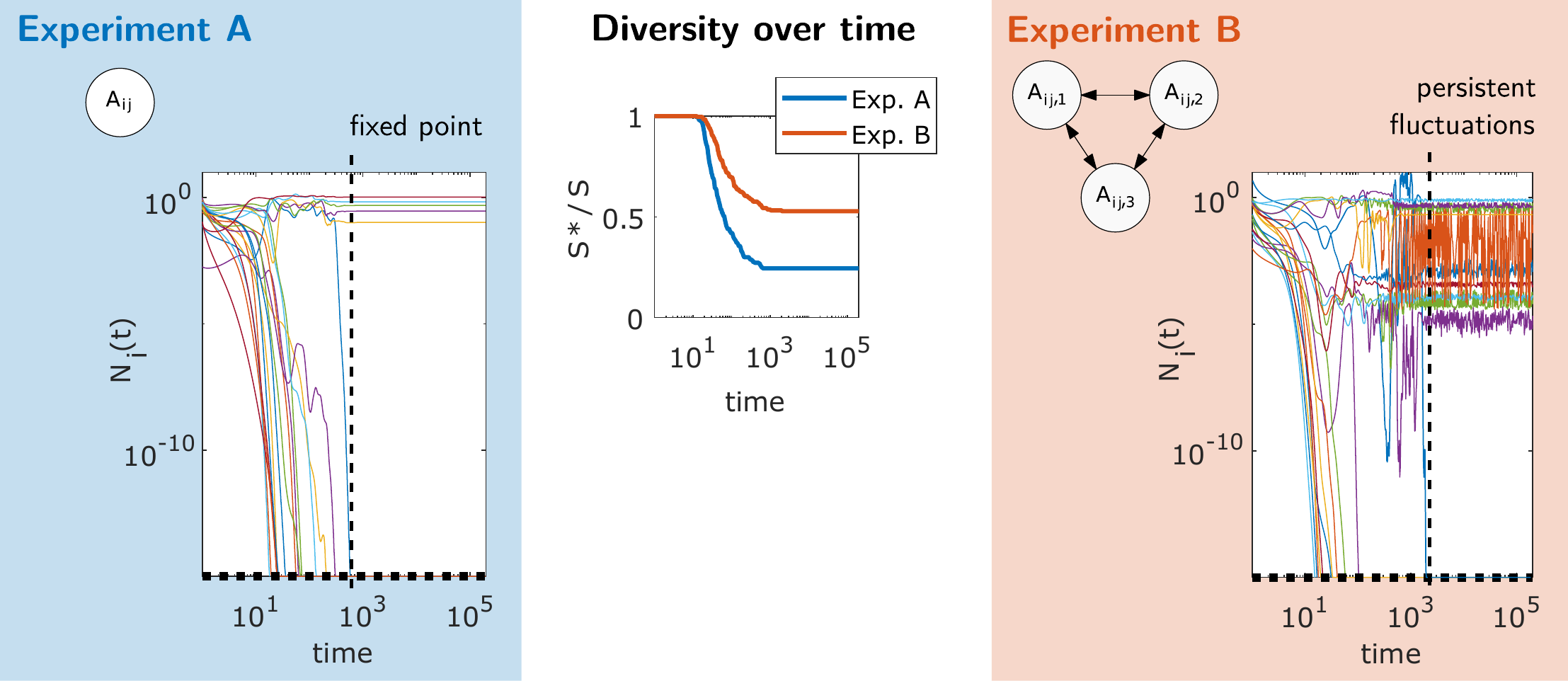} 
\par\end{centering}
\caption{Numerical realization of the proposed experiments, illustrating conditions
that lead to a fixed point or persistent fluctuations. (A) A single
patch (well-mixed community) with an interaction matrix $A_{ij}$.
(B) Multiple patches connected by migration, with slightly different
conditions (e.g. temperature or resources) in each patch, represented
here by location-dependent parameters such as $A_{ij,u}$. In the
right and left panels we show the time evolution of a few representative
species abundances $N_{i}(t)$: Experiment A, with a single patch
($M=1$) reaches a fixed point, while in experiment B a meta-community
with $M=8$ patches reaches a stationary chaotic state ($S=250$).
Middle panel: Fraction of persistent species ($S^{*}$ out of a pool
of $S=250$ species) as a function of time. \label{fig:proposed-experiments}}
\end{figure*}

\medskip{}
 \textbf{Proposed experiments}

In the following, we introduce our results via a set of proposed experiments,
realized in simulations, see Fig. \ref{fig:proposed-experiments}.
These results are later explained in the theoretical analysis. All
parameters for simulations are detailed in Appendix \ref{sec:Appendix:-model-parameters}.

We focus on a meta-community which consists of $M$ patches (well-mixed
systems) connected by migration, and isolated from the external world.
We consider generalized Lotka-Volterra equations for the dynamics
of the abundance $N_{i,u}$ of species $i$ in patch $u$:

\begin{align}
\frac{d}{dt}N_{i,u} & =N_{i,u}\left[B_{i,u}-N_{i,u}-\sum_{j}A_{ij,u}\,N_{j,u}\right]\nonumber \\
 & \ \ \ +\sum_{v}D_{i,uv}\,\left(N_{i,v}-N_{i,u}\right)\ .\label{eq:EOM}
\end{align}
where $A_{ij,u}$ are the interactions coupling the species, $B_{i,u}$
represents the equilibrium abundance in absence of interactions and
migration (known as the carrying capacity), and $D_{i,uv}$ are the
migration rates between patches $u$ and $v$. In addition, an extinction
threshold is implemented as follows: when a species' abundance goes
below a cutoff $N_{c}$ in \emph{all} patches, the species is removed
from the metacommunity and cannot return\footnote{We are interested in the regime where recolonization by migration
between patches is fast compared to the rate of extinction events.
In this regime, we expect (and checked in a few cases) that other
implementations of the cut-off $N_{c}$ will lead to the same qualitative
phenomena. For instance, we implemented patch-wise extinctions when
the abundance goes below the threshold in one particular patch, while
still allowing migrations in.}. This threshold corresponds to the minimum sustainable number of
individuals, hence $1/N_{c}$ sets the scale for the absolute population
size ($P$) of the species. For simplicity, we take $D_{i,uv}=d/\left(M-1\right)$
and $N_{c}$ identical for all species and patches.

The species are assumed to have unstructured interactions (e.g. they
belong to the same trophic level), meaning that $A_{ij,u}$ are sampled
independently\footnote{In the main text we focus on the asymmetric case in which $A_{ij,u}$
and $A_{ji,u}$ are uncorrelated. We show in Appendix \ref{sec:Appendix:gamma_neq_0}
that our results also hold when correlations are present.} and identically for different $\left(i,j\right)$. For a given species
pair, its interactions $A_{ij,u}$ vary somewhat with $u$; this variability
corresponds to small differences in the conditions between the patches
\citep{graham_towards_2018}. In the simulation examples we set all
carrying capacities $B_{i,u}=1$; the phenomena described below are
also found if $B_{i,u}$ vary between patches in addition to, or instead
of the interaction coefficients.

Our proposed experiments, illustrated by dynamical simulations, are
the following:

(A) First, we model a single patch, $M=1$ initially containing $S=250$
species\footnote{Each interaction coefficient is non-zero with probability $c=1/8$,
and the non-zero interactions are Gaussian with $\mean\left(A_{ij,u}\right)=0.3,\,\std\left(A_{ij,u}\right)=0.45$. }. Species go extinct until the system relaxes to a fixed point (stable
equilibrium), see left panel of Fig. \ref{fig:proposed-experiments}.%

(B) We now take $M=8$ patches with the same initial diversity $S=250$
and interaction statistics as in (A). For each pair of interacting
species, $A_{ij,u}$ varies slightly with location $u$, with a correlation
coefficient $\rho=0.95$ between patches. The abundances now fluctuate
without reaching a fixed point, see right panel of Fig. \ref{fig:proposed-experiments}.
At first the diversity decreases as species go extinct, but this process
dramatically slows down, and the diversity is unchanged at times on
the order of $10^{5}$, see middle panel of Fig. \ref{fig:proposed-experiments}.%

\begin{figure}
\begin{centering}
\includegraphics[width=1\columnwidth]{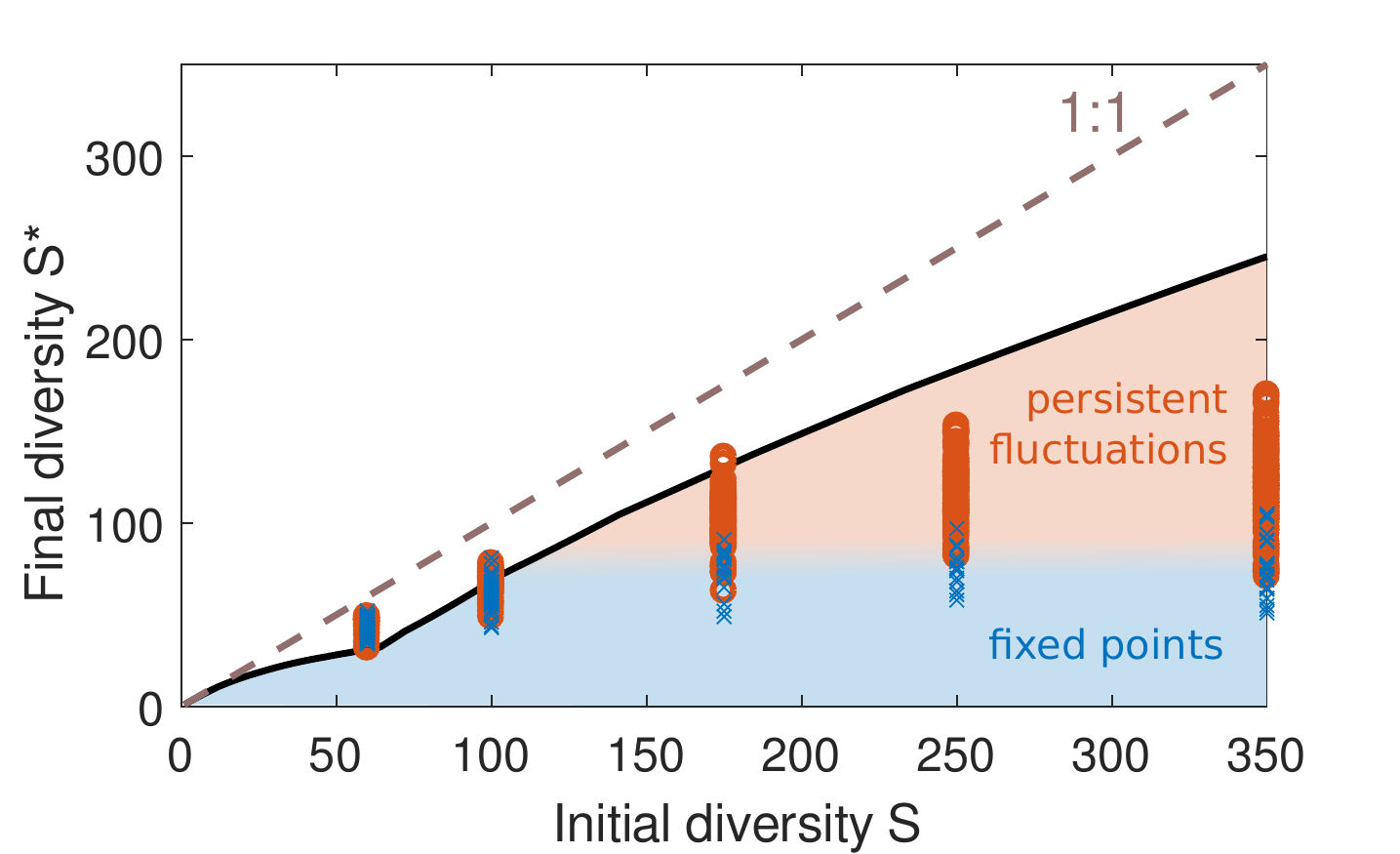} 
\par\end{centering}
\begin{centering}
\includegraphics[width=1\columnwidth]{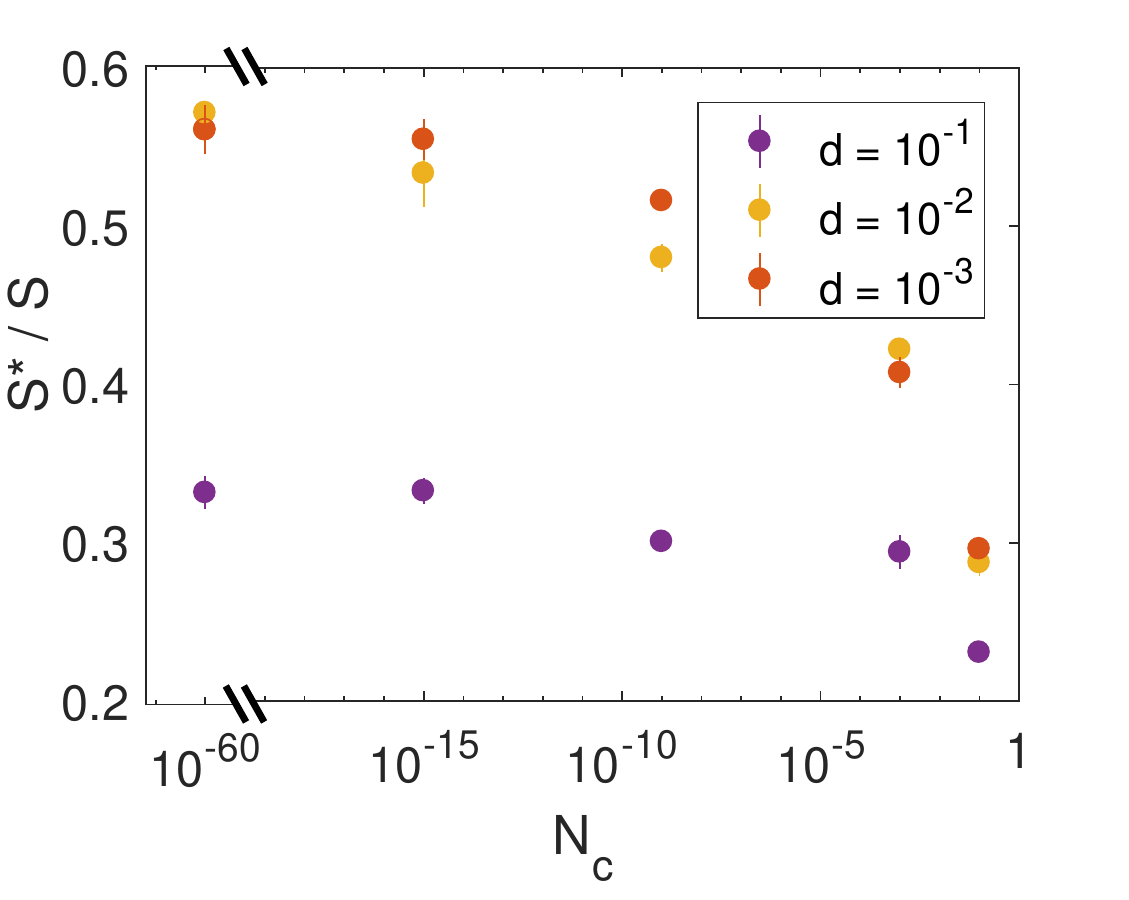} 
\par\end{centering}
\caption{Top: Species diversity at long times, compared to the theoretical
bound obtained in Appendix \ref{sec:Appendix:Diversity-calc} for
large $S$ (solid line). The bound depends on the distribution of
interactions, carrying capacities and initial pool size $S$. Each
symbol represents the state at the end of one simulation run, with
fluctuating states (circles) and fixed points (crosses). States closer
to the theoretical bound (with higher diversity) also exhibit larger
fluctuations and are more difficult to reach due to extinctions in
the transient dynamics (see Fig.\ref{fig:proposed-experiments}).
The dashed line represents full survival ($S^{*}=S$). Bottom: The
final diversity is set by the transient dynamics, which is affected
by factors such as the migration strength and the total population
size ($1/N_{c}$). \label{fig:Theoretical-bounds-on_diversity}}
\end{figure}

Three essential observations emerge from simulating these experiments,
and repeating them for different parameters. First, species diversity
and the strength of endogenous fluctuations are tightly bound, each
contributing to the other's maintenance. Second, as shown in Fig.
\ref{fig:proposed-experiments}, species trajectories first go through
a transient phase where they fluctuate over many orders of magnitude,
causing numerous extinctions which lead to a reduction of variability,
until a fixed point (for $M=1$) or non-equilibrium state (for $M=8$)
with weaker fluctuations is reached. Third, the qualitative difference
between experiments A and B is robust to changes in parameter values.
Changes in $N_{c}$ and $d$ affect only quantitatively the states
that are reached in experiment B, see Fig. \ref{fig:Theoretical-bounds-on_diversity}(top).
For instance, by increasing the population size $P=1/N_{c}$, we can
reach dynamically fluctuating states with higher long-time diversities,
as shown in Fig. \ref{fig:Theoretical-bounds-on_diversity}(bottom).
When the population size is reduced by increasing $N_{c}$, the long-time
diversity decreases, but remains high until $N_{c}\sim10^{-2}-10^{-1}$,
where it decreases dramatically. For example, the diversity shown
in Fig. \ref{fig:proposed-experiments}(right) is $80\%_{\pm13\%}$
higher than that reached for fixed-points at higher $N_{c}$. Similarly,
as long as the migration coefficient is in the range $d\lesssim0.1$
the main qualitative results remain unaltered.%

\medskip{}

\textbf{Theory}

We now aim to understand which conditions allow a fluctuating state
to be reached and maintained without loss of species.\medskip{}

\emph{Dynamical Mean Field Theory}

We build on a powerful theory, known as Dynamical Mean Field Theory
(DMFT) that exactly maps the deterministic meta-community problem
(many species in multiple patches) to a stochastic meta-population
problem (single species in multiple patches). When species traits
and interactions are disordered, e.g. drawn at random from some probability
distributions, all species can be treated as statistically equivalent
\citep{barbier_generic_2018}. We can then describe the whole system
by following the trajectory of a single species, randomly sampled
from the community, and studying its statistics. In the DMFT framework,
the effect of all other species on that single species is encapsulated
by an ``ecological noise'' term generated by their fluctuations.
This is analogous to the use, in physics, of thermal noise to represent
interactions between an open system and its environment. Since species
are statistically equivalent, the properties of this ecological noise
can be self-consistently obtained from the dynamics of the single
species.

While the theory applies to all times \citep{roy_numerical_2019},
as discussed in Appendix \ref{sec:Appendix:-DMFT-equations}, we only
consider here the stationary state\footnote{We will see in the next section that, when endogenous fluctuations
are present, this state is actually metastable, i.e. it is almost
stationary on large but finite time-scales.} reached after a long time, in which extinctions are already rare.
In that state, observables such as the mean abundance are stable over
time, and two-time measures, such as correlation functions between
times $t$ and $t'$, depend only on the difference $t-t'$. This
entails that each species abundance fluctuates with a finite correlation
time, i.e. it tends to return to some constant characteristic value
after a finite time.

The result of this mapping is that the abundance $N_{u}$ of a given
species in patch $u$ undergoes stochastic dynamics, 
\[
\frac{dN_{u}}{dt}=N_{u}\left(N_{u}^{\ast}-N_{u}+\xi_{u}\right)+\sum_{v}D_{uv}\left(N_{v}-N_{u}\right)\ .
\]
This equation models the dynamics of the target species, including
its interactions with other species whose abundances are fluctuating.
The contribution of interactions can be separated into a time-independent
and a time-dependent parts. The time-independent part goes into $N_{u}^{\ast}$,
the characteristic value around which the species abundance will fluctuate
in the patch. It differs between species and between patches, due
to interactions and to environmental preferences modeled by $B_{i,u}$
in Eq. (\ref{eq:EOM}), and follows a multivariate Gaussian distribution.
The time-dependent part is encapsulated in $\xi_{u}\left(t\right)$,
a Gaussian noise with a finite correlation time.

As the quantities $N_{u}^{\ast}$ and $\xi_{u}\left(t\right)$ result
from interaction with other species, which are statistically equivalent
to the target species, one can express their properties from the statistics
of $N_{u}(t)$ itself, as shown in Appendix \ref{sec:Appendix:-DMFT-equations}.
The most important features are that time-dependent Gaussian noise
$\xi_{u}\left(t\right)$ has zero mean and a covariance $C_{\xi}\left(t,t'\right)$,
which is directly related\footnote{It verifies the self consistent equation:
\[
\left\langle \xi_{u}\left(t\right)\xi_{u}\left(t'\right)\right\rangle =\sigma^{2}\left[C_{N,u}\left(t,t'\right)-C_{N,u}^{\infty}\right]
\]
where 
\[
C_{N,u}\left(t,t'\right)=\left\langle N_{u}\left(t\right)N_{u}\left(t'\right)\right\rangle \,\,,
\]
\[
C_{N,u}^{\infty}=\lim_{t-t'\rightarrow\infty}\left\langle N_{u}\left(t\right)N_{u}\left(t'\right)\right\rangle \ .
\]

The average $\langle..\rangle$ over the stochastic process corresponds
to the average over species in the original Lotka-Volterra equations.} to the time auto-correlation $C_{N}\left(t,t'\right)$ of $N_{u}\left(t\right)$
within a patch, and to $\sigma^{2}=cS\thinspace\var\left(A_{ij}\right)$
the variance of interactions rescaled as in \citep{may_will_1972}.
Moreover, the covariance of the $N_{u}^{\ast}$ is fixed by the time
auto-correlation of $N_{u}\left(t\right)$, both within and in-between
patches. In principle, the noise is also correlated between patches,
but this is a small effect in the dynamical regime of interest to
us, see the next section. Note that $C_{\xi}\left(t,t'\right)$ vanishes
when a stable equilibrium is reached.

\begin{figure}
\begin{centering}
\includegraphics[width=1\columnwidth]{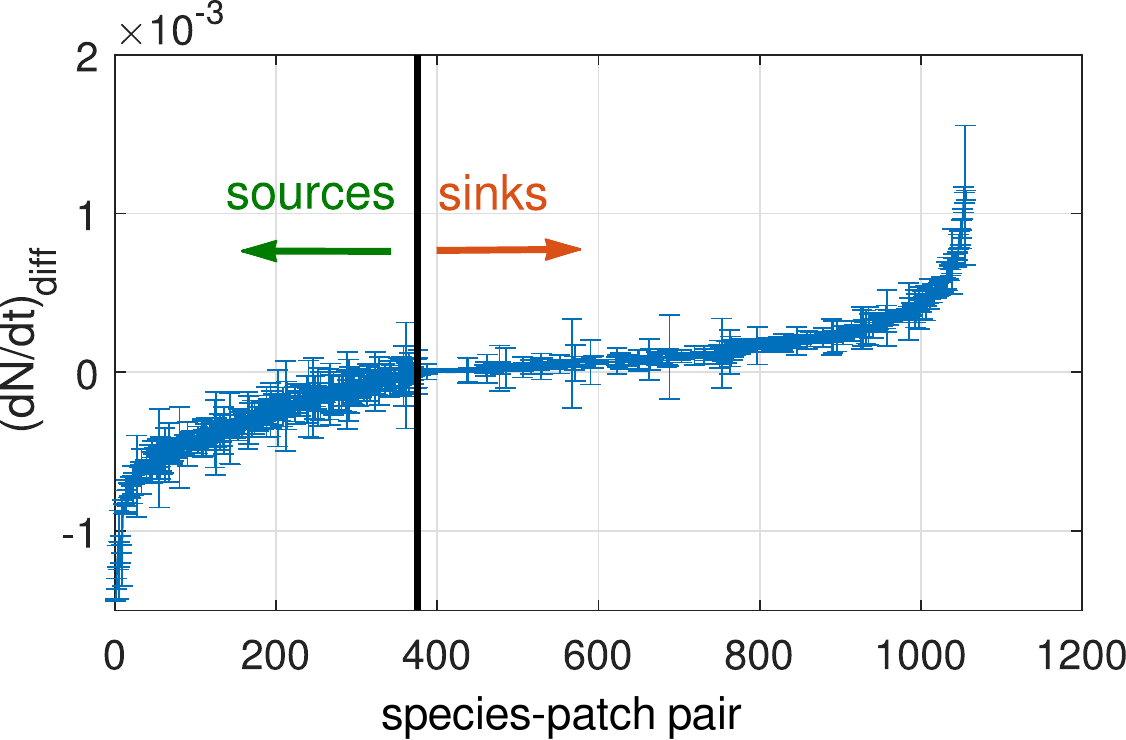} 
\par\end{centering}
\caption{Sources maintain their identity over time. The degree to which a patch
is a source for a given species is measured by $(dN/dt)_{\text{diff}}$,
the contribution of diffusion to the change of $N(t)$, which is negative
for sources and positive for sinks. We show all species-patch pairs
ordered by the average of this quantity over long times, with error
bars giving its standard deviation. For 94\% of sources, and 85\%
of all species-patch pairs, this quantity $(dN/dt)_{\text{diff}}$
retains its sign most of the time, being at least one standard deviation
away from zero.\label{fig:Properties-of-sources} }
\end{figure}

The analysis of the DMFT equations clarifies the main effect of coupling
patches by migration: patches with higher $N_{u}^{\ast}$ tend to
act as sources, i.e. the species most often grows there, and migrates
out to sites where it cannot grow (sinks). We show directly from simulations
of the Lotka-Volterra equations in Fig. \ref{fig:Properties-of-sources}
that species have particular patches which tend to act as sources
consistently over long times. This fact is counter-intuitive, as the
abundances of all species may be fluctuating over orders of magnitude
in any given patch, yet this patch will retain its identity as a source
(or sink) when averaging over long time periods. The variability of
the $N_{u}^{\ast}$s between patches thus leads to an insurance effect,
since it is enough to have one patch acting as a source to avoid extinction
of the species in the others.%

\begin{figure}
\begin{centering}
\includegraphics[width=0.8\columnwidth]{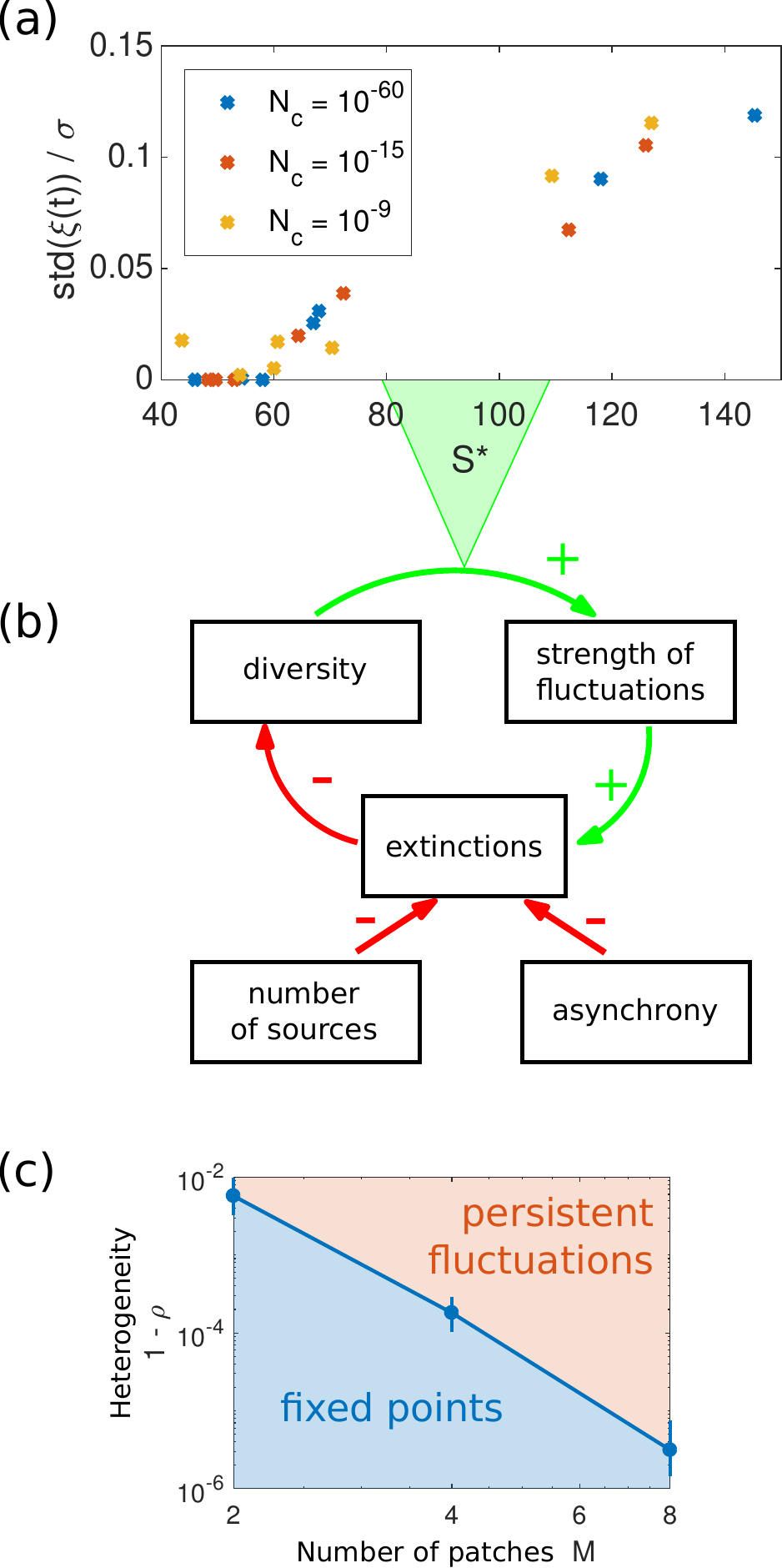} 
\par\end{centering}
\caption{Revisiting the noise-diversity feedback cycle in the light of our
theoretical framework. (a) Quantitative relationship between species
diversity $S^{*}$, i.e. the number of coexisting species, and strength
of fluctuations $\text{{std}\ensuremath{(\xi)}\text{}}$ for $M=8$
(rescaled by interaction heterogeneity $\sigma$). (c) Patch number
$M$ and heterogeneity $1-\rho$ (defined from the correlation coefficient
$\rho$ between interactions $A_{ij,u}$ in different patches $u$)
both contribute to the persistence of endogenous fluctuations by two
means, shown in (b): they create source patches where a given species
will tend to grow (see Fig.\ref{fig:Properties-of-sources}), and
allow the asynchrony of fluctuations in different patches. These two
factors mitigate the likelihood that endogenous fluctuations will
induce species extinctions and cause their own suppression.\label{fig:The-noise-diversity-feedback}}
\end{figure}

\medskip{}
 \emph{Reaching and maintaining a dynamical state}


Let us first consider a single community ($M=1$). For a species to
survive for long periods of time, it follows from DMFT that it must
have positive $N^{\ast}$, or else $N\left(t\right)$ decays exponentially
until the species goes extinct. Even if $N^{\ast}>0$, there is still
a probability (per unit time) of extinction, which depends on $N^{*},\,N_{c}$
and on the strength of the noise $\xi\left(t\right)$. Following extinctions,
a remaining species interacts with fewer fluctuating other species,
causing the strength of the noise to decrease, and with it the probability
for extinction, see feedback loop in Fig. \ref{fig:The-noise-diversity-feedback}.

We can develop an analytical treatment for very small cut-off $N_{c}$
(large population size). In this case there is a large difference
in time-scales between the short-term dynamics induced by endogenous
fluctuations, and the long-term noise-diversity feedback cycle discussed
above. In fact, the extinctions driving this feedback are due to rare
events in which the abundance of species with a positive $N^{*}$
decreases below the (very small) cut-off $N_{c}$. For a species in
an isolated patch ($M=1$), the time-scale for such an event is known
\cite{giles_leigh_average_1981,lande_risks_1993} to be of order of
$\tau\left(1/N_{c}\right)^{a}$ where $\tau$ is a characteristic
time of the endogenous fluctuations, and $a=2N^{*}/W$ is independent\footnote{The expresson of the time scale is analogous to the Arrhenius law
for activated processes in physics and chemistry: in this case, the
counterpart of the energy barrier is $-[N^{*}\ln N_{c}]$ and fluctuation
amplitude $W$ plays the role of the temperature.} of $N_{c}$, with $W$ the amplitude of the endogenous fluctuations,
\[
W\equiv\int dtC_{\xi}\left(t,t'\right)\ .
\]
The important point here is that, although endogenous fluctuations
disappear eventually, there is a clear \textit{separation of time-scales}
between typical endogenous fluctuations, that are fast and lead to
a quasi-stationary dynamical state, and rare extreme fluctuations
that cause extinctions and push the ecosystem into a different state.

While a single community might in principle achieve long-lasting endogenous
fluctuations, this however requires unrealistically large population
sizes and species number, see Appendix \ref{sec:Appendix:-single-patch}.
Migration between multiple patches substantially enhances persistence
due to the spatial insurance effect \citep{loreau_biodiversity_2003}:
species are more unlikely to go extinct because they need to disappear
everywhere at once. The time scale for such an event is $\tau\left(1/N_{c}\right)^{Ma_{\mathrm{eff}}}$
with $a_{\mathrm{eff}}=2N_{\mathrm{eff}}^{\ast}/W$ where $N_{\mathrm{eff}}^{\ast}$
is an effective value for $N^{*}$ of a species across patches, an
expression for which is given in Appendix \ref{sec:Appendix:-extinction-probability}.
This result is thus similar to the one identified above for one patch,
raised to the power $M$. These results assume that $W$ is finite,
and that the noise acting on a species is independent between patches
(asynchrony). Indeed, for moderate values of $D$ and $\rho$ not
too close to one, simulations show that $C_{\xi}\left(t,t'\right)$
is a well-behaved function of $t$ so that $W$ is finite, and the
correlation between patches is found to be very small, see Appendices
\ref{sec:Appendix:-DMFT-equations},\ref{sec:Appendix:-extinction-probability}.

These expressions provide a quantitative description of the feedback
cycle in Fig. \ref{fig:The-noise-diversity-feedback}. Endogenous
fluctuations disappear on the time scale at which species with characteristic
abundance $N_{\mathrm{eff}}^{\ast}$ of order one would go extinct.
We must further account for the vanishing strength of the noise $W$
as species disappear. This is shown in Fig. \ref{fig:The-noise-diversity-feedback}(a),
where the strength of the fluctuations is tightly linked to species
diversity, and is zero at the diversity of fixed points. Hence, extinctions
significantly increase $a_{\mathrm{eff}}$, reducing the chance for
further extinctions. 

\begin{figure}
\begin{centering}
\includegraphics[width=1\columnwidth]{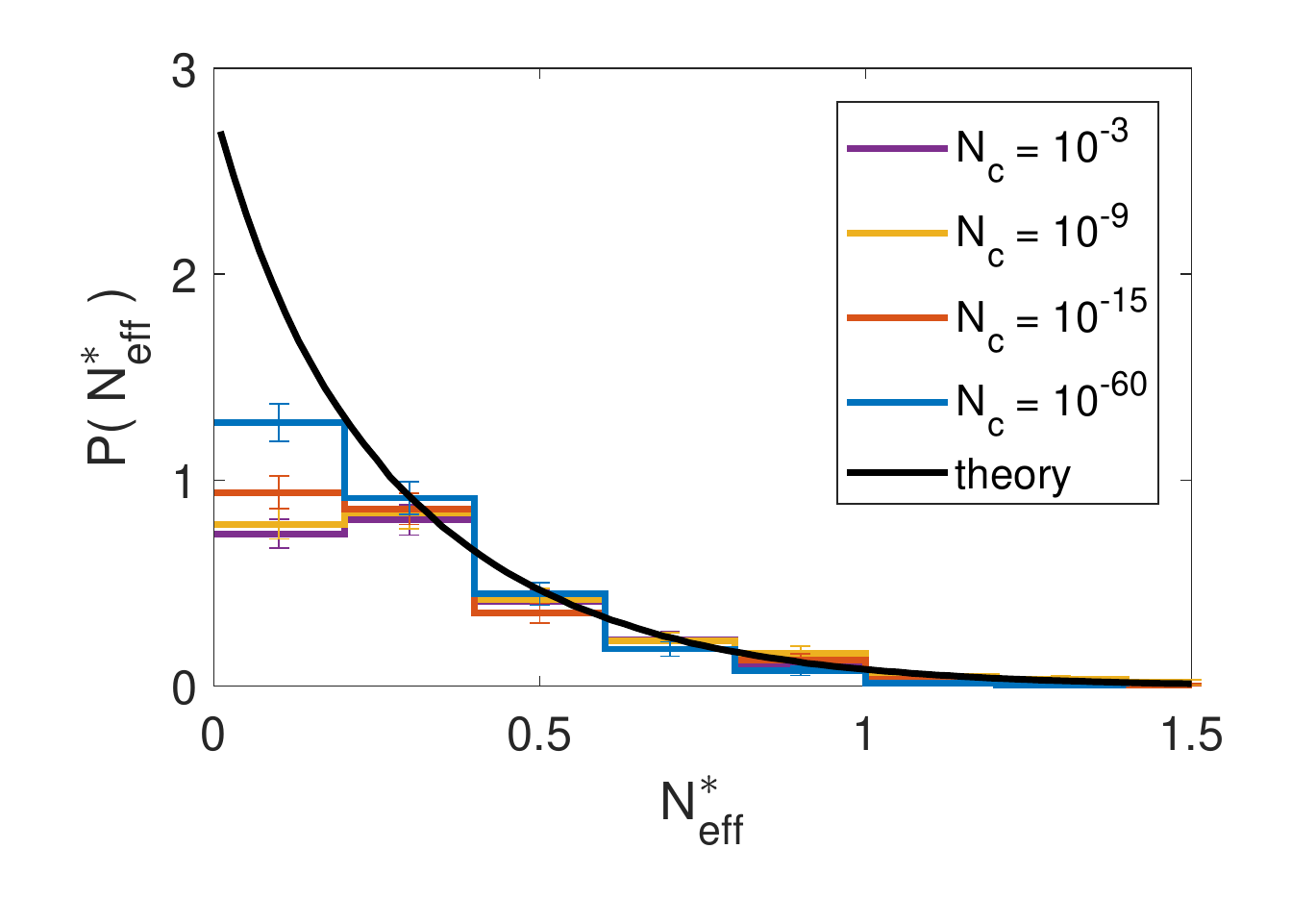} 
\par\end{centering}
\caption{Distributions of the characteristic abundance $N_{\mathrm{eff}}^{*}$
of surviving species, compared with the theoretical prediction for
maximal diversity, showing that the lower diversity in simulations
is mostly due to losing species with lowest $N_{\mathrm{eff}}^{*}$
(leftmost bin). Reducing $N_{c}$ (increasing population size) affects
diversity mainly by allowing these ``rare'' species to persist.\label{fig:Distribution-of-Neff}}
\end{figure}

This picture agrees well with the analytical predictions, which can
be obtained for very small but positive $N_{c}$ and $D$ by DMFT.
These give the distribution of $N_{\mathrm{eff}}^{\ast}$, whose integral
over positive $N_{\mathrm{eff}}^{\ast}$ amounts to the maximal total
diversity at long times. We show in Fig. \ref{fig:Distribution-of-Neff}
that extinct species are generally those that have lower values of
$N_{\mathrm{eff}}^{\ast}$: we compare the analytically predicted
distribution of $N_{\mathrm{eff}}^{\ast}$ to the one observed in
simulations, and see that most missing species had low values of $N_{\mathrm{eff}}^{\ast}$.
The difference becomes smaller for lower $N_{c}$. Due to these differences,
the obtained diversities are lower than the theoretical maximum, see
Fig. \ref{fig:Theoretical-bounds-on_diversity}(top).

As stressed above, the asynchrony of fluctuations in different patches
is crucial: it allows some species to survive with positive characteristic
abundance $N_{i}^{*}$ in at least one of the patches. This leads
to a higher total number of long-term persisting species, decreases
the likelihood of fluctuations to small abundances, and hence increases
the stability of a dynamically fluctuating state. 
If the migration rate $D$ is too strong or $\rho$ very close to
one, dynamics in the different patches synchronize, quickly annulling
the insurance effect. However, minor (few percent) changes in interaction
coefficients or carrying capacities between patches are enough to
maintain this effect, see Fig. \ref{fig:The-noise-diversity-feedback}(bottom);
we don't need to impose coexistence artificially, e.g. by requiring
that every species has at least one refuge (a patch so favorable to
it that it always dominates there). These little variations in the
interaction coefficients are highly plausible, as interaction strength
can vary with many factors, including resource availability \citep{carstensen_beta_2014},
or temperature and its influence on metabolism \citep{kordas_community_2011}.
The heterogeneity $\rho$ required to reach a fluctuating state decreases
with $M$, see Fig. \ref{fig:The-noise-diversity-feedback}(bottom).

In practice, maintaining a dynamical state seems unfeasible for only
one patch, at least for reasonable values of population size $P=1/N_{c}$
and species number $S$, see discussion in Appendix \ref{sec:Appendix:-single-patch}.
Yet the combined effect of the two phenomena described above allows
for very long-lived endogenous fluctuations in metacommunities, already
for $M=2$ patches.

\medskip{}
 \textbf{Conclusions}

Species interactions can give rise to long-lasting fluctuating states,
which both require and allow the maintenance of high species diversity.
This can happen under a wide range of conditions, which we have illustrated
in simulated experiments, and identified through an analytical treatment
based on Dynamical Mean-Field Theory.

While we have drawn parallels with the theory of stability and coexistence
in externally-perturbed ecosystems \cite{loreau_biodiversity_2003,giles_leigh_average_1981,lande_risks_1993,allen_chaos_1993},
our approach also highlights essential differences between environmentally-driven
and endogenous fluctuations. We show that many-species dynamics induce
feedback loops between perturbation and response, and in particular
a tight relationship between fluctuation strength and species diversity,
which are absent from externally-perturbed ecosystems. Moreover, while
similar species can display correlated responses to environmental
stochasticity \cite{loreau_biodiversity_2013}, we expect here that
their trajectories will be starkly different and unpredictable, due
to high-dimensional interactions which lead to complex dynamics. 
The resulting picture from DMFT is that the abundance of any given
species undergoes stochastic dynamics with a finite correlation time.
This means that the trajectory of the species abundance cannot be
predicted after a time that is large compared to the correlation time--a
hallmark of chaos, also found in other models of high-dimensional
systems \cite{sompolinsky_chaos_1988}. Our theory paves the way for
quantitative testing of these fingerprints of diversity-driven fluctuations
in data.

In a counterpoint to classic results \citep{may_will_1972}, we have
shown that, while highly diverse ecosystems are unstable, they might
still persist: extinctions can be avoided and biodiversity maintained,
despite species abundances fluctuating over multiple orders of magnitude.
We do observe a negative feedback loop, in which endogenous fluctuations
cause extinctions, and eventually lead to their own disappearance
as the ecosystem reaches a lower-diversity stable equilibrium. But
this self-suppression of fluctuations can be mitigated by a number
of factors, among which space is particularly important.

In a single well-mixed community, we expect that persistent fluctuations
might not be observed in practice: while theoretically possible, they
may require unrealistic population sizes and species numbers. But
spatial extension and heterogeneity can dramatically reduce these
requirements, in a way that parallels the insurance effect against
exogenous perturbations. When fluctuations are not synchronized across
space, some patches can act as sources, from which failing populations
will be rescued through migration \citep{allen_chaos_1993,loreau_biodiversity_2003}.
Here, we find that the existence of sources is surprisingly robust:
even if there is no location where the environment is favorable to
a given species, source patches can arise from interactions, and endure
for long times despite the large fluctuations in species abundances.
By allowing fluctuations without extinctions, spatial heterogeneity
helps maintain species diversity, and thus the fluctuations themselves.
This result is robust over a wide range of parameters, as it only
calls for moderate values of inter-patch migration: the rate $D$
must be such that, over the typical time scale of abundance fluctuations,
many individuals can migrate out of a patch (allowing recolonization
in the absence of global extinction), while representing only a small
fraction of the population in that patch.

A crucial result is that this condition suffices to ensure that synchronization
between patches is absent, and that the total strength and correlation
time of the noise within patches ($W$ above) remain bounded for finite
populations and finite migration rates between patches. This is in
contrast to alternative scenarios where noise correlations decay slowly
with time \cite{halley_ecology_1996}. This result is non-trivial
for endogenous fluctuations, as the existence of feedbacks (encoded
in the self-consistent equations of the DMFT framework) can potentially
lead to synchronization and long-time correlations in the noise. Yet
we demonstrate that synchrony is avoided, both through direct simulations,
and by building an analytical theory based on these assumptions, whose
predictions match simulations quantitatively.

In conclusion, non-equilibrium fluctuating states might be much more
common than suggested by experiments and theory for well-mixed communities.
And since these fluctuations permit the persistence of more species
than could coexist at equilibrium, we might also expect significantly
higher biodiversity in natural environments.

\medskip{}

\emph{Acknowledgments} - It is a pleasure to thank J.-F. Arnoldi,
J.-P. Bouchaud, C. Cammarota, and M. Loreau for helpful
discussions. 
We thank D. S. Fisher for valuable discussions and sharing the results of his parallel work \cite{dsf}, and in particular for 
key inputs concerning the instability of chaos without migration and dynamical fluctuations in the chaotic phase in the limit of small migration from the mainland. G. Bunin acknowledges support by the Israel Science Foundation
(ISF) Grant no. 773/18. M. Barbier was supported by the TULIP Laboratory
of Excellence (ANR-10-LABX-41) and by the BIOSTASES Advanced Grant,
funded by the European Research Council under the European Union's
Horizon 2020 research and innovation programme (666971). F. Roy acknowledges
support by Capital Fund Management - Fondation pour la Recherche.
G. Biroli was partially supported by the Simons Foundation collaboration
Cracking the Glass Problem (No. 454935).

\bibliographystyle{unsrt}
\bibliography{../10_7_19_only_relevant_files/My_Library}

\begin{thebibliography}{10}

\bibitem{lundberg_population_2000}
Per Lundberg, Esa Ranta, J{\"o}rgen Ripa, and Veijo Kaitala.
\newblock Population variability in space and time.
\newblock {\em Trends in Ecology \& Evolution}, 15(11):460--464, November 2000.

\bibitem{inchausti_relation_2003}
Pablo Inchausti and John Halley.
\newblock On the relation between temporal variability and persistence time in
  animal populations.
\newblock {\em Journal of Animal Ecology}, 72(6):899--908, November 2003.

\bibitem{ellner_chaos_1995}
Stephen Ellner and Peter Turchin.
\newblock Chaos in a noisy world: New methods and evidence from time-series
  analysis.
\newblock {\em The American Naturalist}, 145(3):343--375, 1995.

\bibitem{scheffer_why_2003}
Marten Scheffer, Sergio Rinaldi, Jef Huisman, and Franz~J. Weissing.
\newblock Why plankton communities have no equilibrium: Solutions to the
  paradox.
\newblock {\em Hydrobiologia}, 491(1-3):9--18, January 2003.

\bibitem{may_biological_1974}
Robert~M. May.
\newblock Biological populations with nonoverlapping generations: Stable
  points, stable cycles, and chaos.
\newblock {\em Science}, 186(4164):645--647, 1974.

\bibitem{allen_chaos_1993}
J.~C. Allen, W.~M. Schaffer, and D.~Rosko.
\newblock Chaos reduces species extinction by amplifying local population
  noise.
\newblock {\em Nature}, 364(6434):229--232, July 1993.

\bibitem{may_will_1972}
Robert~M. May.
\newblock Will a large complex system be stable?
\newblock {\em Nature}, 238(5364):413--414, 1972.

\bibitem{berryman_are_1989}
A.A. Berryman and J.A. Millstein.
\newblock Are ecological systems chaotic \textemdash{} {{And}} if not, why not?
\newblock {\em Trends in Ecology \& Evolution}, 4(1):26--28, January 1989.

\bibitem{nisbet_avoiding_1989}
Roger Nisbet, Steve Blythe, Bill Gurney, Hans Metz, Kevin Stokes, Adam
  Lomnicki, and G.S. Mani.
\newblock Avoiding chaos.
\newblock {\em Trends in Ecology \& Evolution}, 4(8):238--240, August 1989.

\bibitem{giles_leigh_average_1981}
Egbert Giles~Leigh.
\newblock The average lifetime of a population in a varying environment.
\newblock {\em Journal of Theoretical Biology}, 90(2):213--239, May 1981.

\bibitem{lande_risks_1993}
Russell Lande.
\newblock Risks of {{Population Extinction}} from {{Demographic}} and
  {{Environmental Stochasticity}} and {{Random Catastrophes}}.
\newblock {\em The American Naturalist}, 142(6):911--927, 1993.

\bibitem{loreau_biodiversity_2003}
Michel Loreau, Nicolas Mouquet, and Andrew Gonzalez.
\newblock Biodiversity as spatial insurance in heterogeneous landscapes.
\newblock {\em Proceedings of the National Academy of Sciences},
  100(22):12765--12770, 2003.

\bibitem{ripa_noise_1996}
J{\"o}rgen Ripa and Per Lundberg.
\newblock Noise colour and the risk of population extinctions.
\newblock {\em Proceedings of the Royal Society of London. Series B: Biological
  Sciences}, 263(1377):1751--1753, 1996.

\bibitem{roy_numerical_2019}
Felix Roy, Giulio Biroli, Guy Bunin, and Chiara Cammarota.
\newblock Numerical implementation of dynamical mean field theory for
  disordered systems: Application to the {{Lotka}}-{{Volterra}} model of
  ecosystems.
\newblock {\em Journal of Physics A: Mathematical and Theoretical}, 2019.

\bibitem{gravel_stability_2016}
Dominique Gravel, Fran{\c c}ois Massol, and Mathew~A. Leibold.
\newblock Stability and complexity in model meta-ecosystems.
\newblock {\em Nature Communications}, 7:12457, August 2016.

\bibitem{graham_towards_2018}
Catherine~H. Graham and Ben~G. Weinstein.
\newblock Towards a predictive model of species interaction beta diversity.
\newblock {\em Ecology Letters}, 21(9):1299--1310, September 2018.

\bibitem{barbier_generic_2018}
Matthieu Barbier, Jean-Fran{\c c}ois Arnoldi, Guy Bunin, and Michel Loreau.
\newblock Generic assembly patterns in complex ecological communities.
\newblock {\em Proceedings of the National Academy of Sciences},
  115(9):2156--2161, February 2018.

\bibitem{carstensen_beta_2014}
Daniel~W. Carstensen, Malena Sabatino, Kristian Tr{\o}jelsgaard, and Leonor
  Patricia~C. Morellato.
\newblock Beta {{Diversity}} of {{Plant}}-{{Pollinator Networks}} and the
  {{Spatial Turnover}} of {{Pairwise Interactions}}.
\newblock {\em PLOS ONE}, 9(11):e112903, November 2014.

\bibitem{kordas_community_2011}
Rebecca~L. Kordas, Christopher~D.G. Harley, and Mary~I. O'Connor.
\newblock Community ecology in a warming world: {{The}} influence of
  temperature on interspecific interactions in marine systems.
\newblock {\em Journal of Experimental Marine Biology and Ecology},
  400(1-2):218--226, April 2011.

\bibitem{loreau_biodiversity_2013}
Michel Loreau and Claire de~Mazancourt.
\newblock Biodiversity and ecosystem stability: A synthesis of underlying
  mechanisms.
\newblock {\em Ecology Letters}, 16(s1):106--115, 2013.

\bibitem{sompolinsky_chaos_1988}
Haim Sompolinsky, Andrea Crisanti, and Hans-Jurgen Sommers.
\newblock Chaos in random neural networks.
\newblock {\em Physical Review Letters}, 61(3):259, 1988.

\bibitem{halley_ecology_1996}
John~M. Halley.
\newblock Ecology, evolution and 1f-noise.
\newblock {\em Trends in Ecology \& Evolution}, 11(1):33--37, January 1996.

\bibitem{mezard_bethe_2001}
Marc M{\'e}zard and Giorgio Parisi.
\newblock The {{Bethe}} lattice spin glass revisited.
\newblock {\em The European Physical Journal B-Condensed Matter and Complex
  Systems}, 20(2):217--233, 2001.

\bibitem{ovaskainen_stochastic_2010}
Otso Ovaskainen and Baruch Meerson.
\newblock Stochastic models of population extinction.
\newblock {\em Trends in Ecology \& Evolution}, 25(11):643--652, November 2010.

\bibitem{kamenev_how_2008}
Alex Kamenev, Baruch Meerson, and Boris Shklovskii.
\newblock How {{Colored Environmental Noise Affects Population Extinction}}.
\newblock {\em Physical Review Letters}, 101(26):268103, December 2008.

\bibitem{bunin_ecological_2017}
Guy Bunin.
\newblock Ecological communities with {{Lotka}}-{{Volterra}} dynamics.
\newblock {\em Physical Review E}, 95(4), April 2017.

\bibitem{dsf}
M.T. Pearce, A. Agarwala, and D. S. Fisher. 
\newblock Stabilization of extensive fine-scale diversity by spatio-temporal chaos. 
\newblock {\em bioRxiv} (2019): 736215.

\end{thebibliography}
\newpage{}

\appendix

\section{Model parameters used in simulations\label{sec:Appendix:-model-parameters}}

For convenient reference, this Appendix includes the parameters for
all simulations. The model is given in Eq. (\ref{eq:EOM}). All $B_{i,u}=1$
and all $D_{i,uv}=d/\left(M-1\right)$. The $A_{ij,u}$ are independent
for different $\left(i,j\right)$ pairs (except in Appendix \ref{fig:gam_neq_0}).

In Fig. \ref{fig:proposed-experiments}, the probability of $A_{ij,u}$
to be non-zero is $c=1/8$, and the non-zero elements are sampled
from a normal distribution with $\mean\left(A_{ij,u}\right)=0.3,\,\std\left(A_{ij,u}\right)=0.45$.
The same elements $A_{ij,u}$ are non-zero across all patches $u$.
The correlation coefficient between non-zero $A_{ij,u}$ in different
patches is $\rho=\corr\left[A_{ij,u},\,A_{ij,v}\right]=0.95$ for
$u\neq v$. (The correlation is $0.964$ when interactions with $A_{ij,u}=0$
are also counted.) The initial (pool) diversity is $S=250$. In Fig.
\ref{fig:proposed-experiments}(A), $M=1$. In Fig. \ref{fig:proposed-experiments}(B),
$M=8$ patches and $d=10^{-3}$. The cutoff is $N_{c}=10^{-15}$.

Fig. \ref{fig:Theoretical-bounds-on_diversity}(bottom), uses the
same parameters as Fig. \ref{fig:proposed-experiments}, but with
a range of values for $d,S$ and $N_{c}$.

Fig. \ref{fig:Properties-of-sources} uses the runs shown in Fig.
\ref{fig:proposed-experiments}(B). Standard deviation and mean are
estimated from 1601 time points during the time period $t=[10^{4},2\cdot10^{5}]$.
Fig. \ref{fig:Distribution-of-Neff} uses multiple runs, with the
same parameters as \ref{fig:proposed-experiments}, except for $d=10^{-4}$
and the values of $N_{c}$ that are detailed in the figure legend.

Fig. \ref{fig:The-noise-diversity-feedback}(top) uses the same parameters
as \ref{fig:Distribution-of-Neff}. Fig. \ref{fig:The-noise-diversity-feedback}(bottom),
shows the line where half of the runs are fixed points, and half continue
to fluctuate until $t=2\cdot10^{5}$. It uses same parameters as \ref{fig:Distribution-of-Neff},
except with $D=d/\left(M-1\right)=10^{-4}$.

In Fig. \ref{fig:The-noise-diversity-feedback}(top), the size of
the fluctuations are calculated from $\var\left(\xi_{u}\right)=\left\langle \xi_{u}^{2}\left(t\right)\right\rangle =\sigma^{2}C_{N,u}\left(t,t\right)$,
with $C_{N,u}\left(t,t\right)=\left\langle N_{u}^{2}\left(t\right)\right\rangle -\lim_{t-t'\rightarrow\infty}\left\langle N_{u}\left(t\right)N_{u}\left(t'\right)\right\rangle $.
For more details on the averaging, see Appendix \ref{sec:Appendix:-DMFT-equations},
Fig. \ref{fig:TTI_xi}.

\section{DMFT equations\label{sec:Appendix:-DMFT-equations}}

In this section, we present the full DMFT equations, and explain how
they can be reduced to the steady-state equations quoted in the main
text.

We consider as a starting point equation Eq. (\ref{eq:EOM}). For
the sake of clarity, we derive DMFT under simplifying assumptions,
but the result is much more robust and could be applied to different
ecology models as well as real data \citep{barbier_generic_2018}.
DMFT for ecological models has a double valency analogous to the one
of mean-field theories in physics: it is at the same time an exact
theory for some simple models, and a powerful approximation largely
applicable to a broad range of systems. For the sake of clarity, the
derivation assumes a fully connected model (all interactions are non-zero),
but the results hold for any connectivity $C$ as long as $C\gg1$,
see remark at the end of this Appendix.

The assumptions which make DMFT exact are the following: all constants
$N_{i,u}(0)$, $B_{i,u}$, $D_{i,uv}$ and $A_{ij,u}$ are random
variables, sampled from known distributions. More precisely: 
\begin{itemize}
\item in each patch $u$ and for all species $i$, the parameters $X^{u}=\lbrace N_{i,u}(0),\,B_{i,u},\,D_{i,uv}\rbrace_{i=1}^{S}$
are drawn from a probability distribution $\mathbb{P}$ which is a
product measure $\mathbb{P}_{u}(X^{u})=\prod_{i=1}^{S}\mathbb{P}(X_{i}^{u})$; 
\item the interaction matrix can be decomposed as $A_{ij,u}=\mu/S+\sigma/\sqrt{S}\;a_{ij,u}$.
$a_{ij,u}$ are standard random variables with mean zero, variance
one, and correlation: 
\[
\mathbb{E}\left[a_{ij,u}\,a_{kl,v}\right]=\delta_{ik}\,\delta_{jl}\,\rho_{uv}
\]

where we used the Kronecker symbol $\delta_{ik}$, and $\rho_{uv}=\rho+(1-\rho)\delta_{uv}$
is a uniform correlation $\rho$ between patches. 
\end{itemize}
With these conventions, we rewrite Eq. (\ref{eq:EOM}) in the following
way:

\begin{align*}
\frac{d}{dt}N_{i,u} & =N_{i,u}\left[B_{i,u}-N_{i,u}-\mu\,m_{u}(t)+\eta_{i,u}(t)\right]\\
 & \ \ \ +\sum_{v}D_{i,uv}\left(N_{i,v}-N_{i,u}\right)\ 
\end{align*}

where $m_{u}(t)=S^{-1}\,\sum_{i=1}^{S}N_{i,u}(t)$ is the mean abundance
in patch $u$, and $\eta_{i,u}(t)=-\sigma S^{-1/2}\,\sum_{j=1}^{S}a_{ij,u}\,N_{j,u}(t)$
is a species-and-patch-dependent noise.

The DMFT equation can be obtained by following Ref. \citep{roy_numerical_2019}:
in the large-$S$ limit, it can be shown that the statistics of this
multi-species deterministic process corresponds to the following one-species
stochastic process, for each patch.

\begin{align}
\frac{d}{dt}N_{u} & =N_{u}\left[B_{u}-N_{u}-\mu\,m_{u}(t)+\eta_{u}(t)\right]\nonumber \\
 & \ \ \ +\sum_{v}D_{uv}\left(N_{v}-N_{u}\right)\ \label{eq:fullDMFT}
\end{align}

where $\lbrace N_{u}(0),\,B_{u},\,D_{uv}\rbrace$ are sampled from
the distribution $\mathbb{P}(X^{u})$, $m_{u}(t)$ is a deterministic
function, and $\eta_{u}(t)$ is a zero-mean Gaussian noise. The variability
from one species to another becomes in the DMFT setting the randomness
contained in $\lbrace N_{u}(0),\,B_{u},\,D_{uv}\rbrace$ and $\eta_{u}(t)$.

To make this point crystal clear, let us introduce two different averages: 
\begin{itemize}
\item $\overline{Y}$ averages over the stochastic process in Eq. (\ref{eq:fullDMFT}):
over the stochastic noise $\eta_{u}$ and over the distribution $\mathbb{P}(X^{u})$; 
\item $\mathbb{E}_{S}(Y)$ denotes the statistical average over the deterministic
multi-species system. $\mathbb{E}_{S}(Y)=\sum_{i=1}^{S}Y_{i}$, and
therefore also includes sampling of $X_{i}^{u}$. 
\end{itemize}
DMFT represents in terms of a stochastic process the deterministic
dynamical system governing the dynamics of the $S$ species in the
ecosystem. In consequence, averages over the stochastic process coincide
with average over species: for a given observable $Y$: $\overline{Y}=\lim_{S\to\infty}\mathbb{E}_{S}(Y)$.
This is analogous to the representation of the environment of an open
physical system in terms of thermal noise, as it is done e.g. in the
case of the Langevin equation.

The second important aspect of DMFT is \textit{self-consistency}.
This is related to the fact that the noise is induced by the dynamics
of the species themselves, so its properties can be obtained from
dynamical averages:

\[
\left\{ \begin{aligned} & m_{u}(t)=\overline{N_{u}(t)}\\
 & \langle\eta_{u}(t)\eta_{v}(t')\rangle=\sigma^{2}\,\rho_{uv}\;\overline{N_{u}(t)\,N_{v}(t')}
\end{aligned}
\right.
\]

where we used a last average $\langle\cdot\rangle$ over the stochastic
noise only, in order to define its covariance. Henceforth we use the
notation $C_{uv}^{N}(t,t')=\rho_{uv}\,\overline{N_{u}(t)\,N_{v}(t')}$.

We now show how DMFT equations simplify for a time-translationally-invariant
state of the system, which is in general reached after some transient
time. 
In this state, all one-time observables become constant in time, and
two-time observables become functions of the time difference only.

\[
\left\{ \begin{aligned} & m_{u}=\overline{N_{u}(t)}\\
 & C_{uv}^{N}(t,t')=C_{uv}^{N}(t-t')
\end{aligned}
\right.
\]

The correlation $C_{uv}^{N}(t-t')$ decays at large time differences
to a non-zero constant, leading to a static contribution to the noise
term. In order to disentangle the static part and the time-fluctuating
part of the noise, we perform the decomposition $\eta_{u}(t)=z_{u}+\xi_{u}(t)$
such that $z_{u}$ and $\xi_{u}(t)$ are independent zero-mean Gaussian
variables and processes verifying: 
\[
\langle z_{u}z_{v}\rangle=\sigma^{2}\,\lim_{t-t'\rightarrow\infty}C_{uv}^{N}(t-t')
\]
and subsequently $\langle\xi_{u}(t)\xi_{v}(t')\rangle=\sigma^{2}\,C_{uv}^{N}(t-t')-\sigma^{2}\,\lim_{t-t'\rightarrow\infty}C_{uv}^{N}(t-t')$
which vanishes for $t-t'\rightarrow\infty$.

Substituting this decomposition into Eq. (\ref{eq:fullDMFT}), we
obtain:

\begin{align}
\frac{d}{dt}N_{u} & =N_{u}\left[N_{u}^{*}-N_{u}+\xi_{u}(t)\right]\nonumber \\
 & \ \ \ +\sum_{v}D_{uv}\left(N_{v}-N_{u}\right)\ 
\end{align}

where $N_{u}^{*}=1-\mu\,m_{u}+z_{u}$ is a Gaussian variable, whose
statistics is described in Appendix \ref{sec:Appendix:Diversity-calc}.
We checked numerically that for small migration $D$, the noise is
only correlated between patches through its static part: for $u\neq v$,
$\xi_{u}(t)\,\xi_{v}(t')\ll z_{u}\,z_{v}$, as presented in Fig. \ref{fig:TTI_xi}.
In this case, we can write the self-consistent closure as follows:

\[
\left\{ \begin{aligned} & m_{u}=\lim_{t'\gg1}\overline{N_{u}(t')}\\
 & \langle z_{u}z_{v}\rangle=\sigma^{2}\,\lim_{t\gg t'\gg1}C_{uv}^{N}(t,t')\\
 & \langle\xi_{u}(t)\xi_{v}(t')\rangle=\delta_{uv}\,\sigma^{2}\,\left[C_{uv}^{N}(t,t')-\lim_{t\gg t'\gg1}C_{uv}^{N}(t,t')\right]
\end{aligned}
\right.
\]

\begin{center}
\begin{figure}
\centering{}\includegraphics[width=0.5\columnwidth]{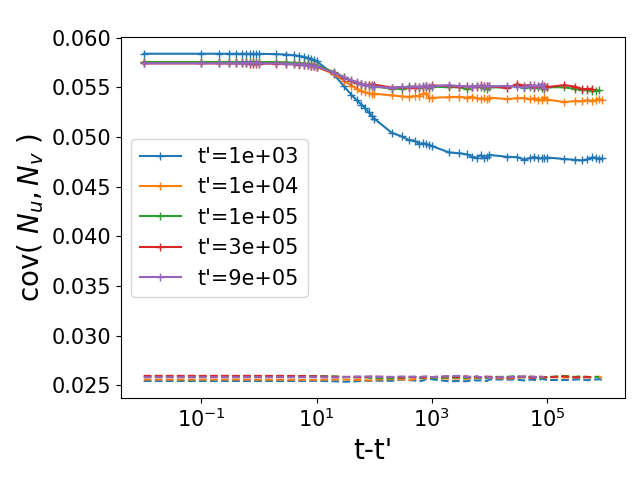}\includegraphics[width=0.5\columnwidth]{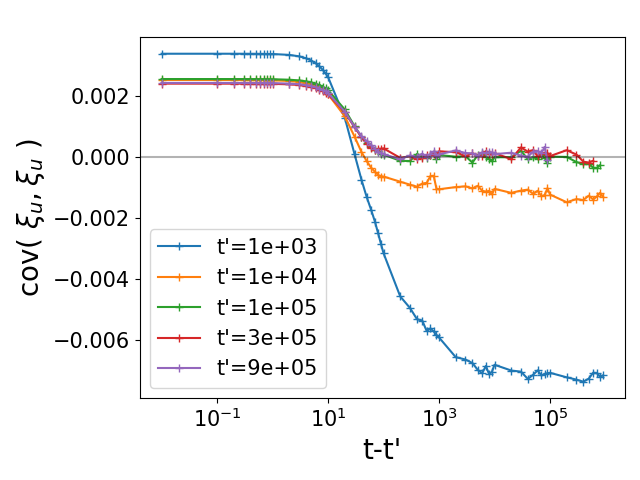}
\caption{Covariance of the abundances in distinct patches. We use the general
notation $\mathrm{cov}(Y_{u},Y_{v})=\mathbb{E}_{S}[Y_{u}^{c}(t)\,Y_{v}^{c}(t')]$
and $Y_{u}^{c}(t)=Y_{u}(t)-\mathbb{E}_{S}[Y_{u}(t)]$. Left: In full
lines we show the abundance covariance within a patch $u=v$, and
across patches $u\protect\neq v$ in dotted lines. The correlation
in abundances across patches is mainly static: dotted lines are reasonably
flat. In other words, the correlation of $\xi_{u}$ with $\xi_{v}$
for $u\protect\neq v$ is very small. Right: The covariance in $\xi$
is shown to reach a TTI state. It only depends on $t-t'$ after $t'=10^{5}$:
the colored curves collapse. In this data, 100 distinct simulations
were averaged, with parameters $(S,\mu,\sigma\,|\,M,\rho,d,N_{c})=(400,10,2\,|\,8,0.95,10^{-10},10^{-15})$.}
\label{fig:TTI_xi} 
\end{figure}
\par\end{center}

As explained above, DMFT can be implemented as an approximation for
a large variety of systems. In this case one has to infer the average
$\mu$, the standard deviation $\sigma$ of interactions, and the
distribution $\mathbb{P}(X^{u})$ from the data (we remind that $X^{u}=\lbrace N_{u}(0),\,B_{u},\,D_{uv}\rbrace$)
and use them as an input to define an effective model. The generalization
to patch-dependent cumulants $\mu_{u}$ and $\sigma_{u}$ is quite
straightforward. So is the generalization to patch-dependent correlation
$\rho_{uv}$.

We have derived DMFT for a completely connected set of interactions
$A_{ij}$. A different way to obtain DMFT is considering a finite
connectivity network of interactions $A_{ij}$, e.g. the one produced
by a Erdos-Renyi random graph with average connectivity per species
$C$ or a regular random graph with connectivity $C$. In these cases,
for each link $ij$ one generates a random variable with average $\mu/C$
and variance $\sigma^{2}/C$ and set it to $A_{ij}$. In the large
connectivity limit, $C\rightarrow\infty$, each species interacts
with a very large number of species and one can replace the deterministic
interaction with an effective stochastic noise, as done for a completely
connected lattice. Although the resulting DMFT equations are the same,
the two cases are quite different: in the former a species interact
with $C\ll S$ species whereas in the latter a species interacts with
$C=S$ species. The equivalence of DMFT for completely connected lattices
and finite connectivity ones in the $C\rightarrow\infty$ limit has
been thoroughly studied in physics of disordered systems in the last
twenty years \citep{mezard_bethe_2001}.

\section{Extinction probability of a species\label{sec:Appendix:-extinction-probability}}

Here the probability of extinction of a species is presented, at the
limit $N_{c}\ll D\ll1$. More specifically, we assume that $N_{c}$
is small compared to the typical fluctuations of the abundances. In
addition, in simulations we see that it is reasonable to assume complete
lack of synchrony, namely that the noise $\xi_{u}$ is uncorrelated
between different patches, see Appendix \ref{sec:Appendix:-DMFT-equations},
Fig. \ref{fig:TTI_xi}. We will therefore assume that in the following
calculation. Finally, we assume that for at least one patch, $N_{u}^{*}>0$,
otherwise the species goes quickly extinct.

Within DMFT, the problem thus becomes ones of calculating the extinction
probability of a meta-population (single species), under environmental
fluctuations, that are uncorrelated between the different patches.
We only present the result here; a full account will be given elsewhere.

Let $x_{u}\equiv\ln N_{u}$. The equations of the DMFT, Eq. (\ref{eq:fullDMFT}),
become 
\begin{equation}
\partial_{t}x_{u}=N_{u}^{\ast}-e^{x_{u}}-\sigma\xi_{u}+D\sum_{v}\left(e^{x_{v}-x_{u}}-1\right)\ .\label{eq:DMFT_x_vars}
\end{equation}
Here $D=d/\left(M-1\right)$. We look for a rare realization of $\left\{ \xi_{u}\right\} $
that makes all the $x_{u}$ reach $x_{c}=\ln N_{c}$, in the case
when the cut-off is low, $x_{c}\rightarrow-\infty$. The calculation
proceeds within the framework of large-deviation theory \cite{ovaskainen_stochastic_2010}.
First, one defines the ``action'' 
\begin{equation}
J=\frac{1}{2W}\int^{t_{f}}dt\sum_{u=1}^{M}\xi_{u}^{2}\ ,\label{eq:DMFT_LDF_action}
\end{equation}
with $\xi_{u}$ substituted with its value from Eq. (\ref{eq:DMFT_x_vars}),
and $W$ defined as in the main text. Here we approximated the noise
correlations by white noise, which is justified here as the extinction
event takes a time which is long compared to the correlation time.
We assume that $D$ is small.

Then the mean time to the occurrence of such an event scales as $P\sim e^{J_{min}}$
with $J_{min}$ the action $J$ minimized over all population trajectories
$\left\{ x_{u}\left(t\right)\right\} _{u=1..M}$ that start at $t\rightarrow-\infty$
at the typical value of $x_{u}$, obtained by the zero-noise fixed
point of Eq. (\ref{eq:DMFT_x_vars}), and terminate at $t_{f}$ at
$x_{c}=\ln N_{c}$ .

We first describe the result for $M=1$. In this case there is only
one patch, $u=1$, with $N_{1}^{\ast}$. If $N_{1}^{\ast}<0$ the
species is extinct. On the other hand, if $N_{1}^{\ast}>0$, then
we obtain the known result \cite{lande_risks_1993,kamenev_how_2008}
\[
J_{\min}=\frac{2x_{c}}{W}N_{1}^{\ast}\ .
\]
The result for all $M$ is a generalization of this result, of the
form 
\[
J_{\min}=\frac{2x_{c}}{W}MN_{\mathrm{eff}}^{\ast}\ .
\]

To describe the calculation of $N_{\mathrm{eff}}^{\ast}$, order the
patches so that $N_{1}^{\ast}\ge N_{2}^{\ast}\ge..\ge N_{M}^{*}$.
Then there exists $1\le m\le M$ such that

\[
w\equiv-\sqrt{\frac{1}{m}\sum_{u=1}^{m}\left(N_{u}^{\ast}\right)^{2}}\ ,
\]
and where $w$ satisfies: $w\le N_{u}^{\ast}$ for all $u\le m$,
and $w>N_{u}^{\ast}$ for all $u>m$. Such a partition can be shown
to always exist. Then 
\[
N_{\mathrm{eff}}^{\ast}\equiv\frac{1}{4M}\sum_{\left\{ u\right\} _{+}}\frac{\left(N_{u}^{\ast}-w\right)^{2}}{w}\ .
\]
The derivation will be given elsewhere. We illustrate the result by
considering two cases. First, in the $M=1$ example, since $N_{1}^{\ast}>0$,
the partition is $\left\{ u\right\} _{+}=\left\{ 1\right\} $ and
$\left\{ u\right\} _{-}$ the empty set. Indeed, this gives $w=-N_{1}^{\ast}$,
so $w\le N_{1}^{*}$. Then $N_{\mathrm{eff}}^{\ast}=N_{1}^{\ast}$
and $J_{\min}=\frac{2x_{c}}{W}N_{1}^{\ast}$, so the result for $M=1$
is reproduced. Another simple case is when there are $M$ patches
with identical carrying capacities $N_{u=1..M}^{\ast}=N^{\ast}$.
Here $\left\{ u\right\} _{+}=\left\{ 1,..,M\right\} $ and $w=-N^{\ast}$.
Then $N_{\mathrm{eff}}^{\ast}=N^{\ast}$, and $J_{\min}=\frac{2x_{c}}{W}MN_{\mathrm{eff}}^{\ast}=\frac{2x_{c}}{W}MN^{\ast}$.
This result is intuitively clear: to go extinct, the species must
go extinct in all patches at once, so the probability is $P\sim\exp\left(\frac{x_{c}}{W}MN^{\ast}\right)\sim\left(P_{1}\right)^{M}$,
where $P_{1}$ is the result for $M=1$.

\section{Diversity and stability at low migration rates\label{sec:Appendix:Diversity-calc}}

We use notations from Appendix \ref{sec:Appendix:-DMFT-equations}.
Within the time-translational-invariant state: 
\begin{align*}
\frac{1}{N_{u}}\frac{dN_{u}}{dt} & =N_{u}^{*}-N_{u}+\xi_{u}\left(t\right)+\sum_{v\sim u}D_{uv}\left(\frac{N_{v}}{N_{u}}-1\right)
\end{align*}
Consider the case of low migration, $D\rightarrow0^{+}$. We now develop
a theory assuming that the amplitude of the endogenous fluctuations,
\[
W\equiv\int dt\:C_{\xi}\left(t,t'\right)\ ,
\]
remains finite in the limit $D\rightarrow0^{+}$. Assume the species
survives, i.e. there is at least one patch with $N_{u}^{*}>0$. If
$N_{u}^{*}<0$ then $N_{u}=O\left(D\right)$. If $N_{u}^{*}>0$ then
$N_{u}=O\left(1\right)$ and therefore $\sum_{v\sim u}D_{uv}\left(\frac{N_{v}}{N_{u}}-1\right)=O\left(D\right)$.
Taking the time average of the above equation 
\[
0=\overline{\frac{1}{N_{u}}\frac{dN_{u}}{dt}}=N_{u}^{*}-\overline{N_{u}}+O\left(D\right)
\]
and therefore $\overline{N_{u}}=N_{u}^{*}+O\left(D\right)$.

The previous arguments lead to the conclusion that in the $D\rightarrow0^{+}$
limit $\overline{N_{u}}=N_{u}^{*}$ if $N_{u}^{*}>0$ and is equal
to zero otherwise.%
{} In the following we provide more detail more this argument and its
possible limitations. For this last equality to be valid, we need
that $\sum_{v\sim u}\left(\frac{N_{v}}{N_{u}}-1\right)$ will be finite,
so that $D\sum_{v\sim u}\left(\frac{N_{v}}{N_{u}}-1\right)$ will
indeed be small. This might break if $N_{u}$ can be small while some
other $N_{v}$ remains $O\left(1\right)$. An estimate for that proceeds
by noting that the carrying capacity of patch $u$ in the presence
of other patches is larger or equal to $N_{u}^{*}-MD\simeq N_{u}^{*}$,
its carrying capacity alone. If patch $u$ fluctuates alone, then
\[
\frac{dx_{u}}{dt}=N_{u}^{*}+\xi\left(t\right)\Rightarrow P\left(x\right)\sim e^{\frac{2N^{*}x}{\sigma^{2}W}}
\]
This gives for $\overline{1/N_{u}}$

\begin{align*}
\overline{e^{-x_{u}}} & \sim\frac{\int_{-\infty}^{0}e^{x\left(\frac{N_{u}^{*}}{W}-1\right)}dx}{\int_{-\infty}^{0}e^{x\frac{N_{u}^{*}}{W}}dx}=\frac{\frac{N_{u}^{*}}{W}+1}{\frac{N_{u}^{*}}{W}}=1+\frac{W}{N_{u}^{*}}
\end{align*}
For any given $N_{u}^{*}$ this is finite. It diverges as $N_{u}^{*}\rightarrow0$.
Therefore the migration term is negligible only if $\frac{DW}{1-D}\simeq DW\ll N_{u}^{*}$.
(Note that migration itself would limit $N_{u}$ going below much
below $DN_{v}$, which would make this term smaller.) The main approximation
(or limitation) of our approach is the assumption that $W$ remains
finite in the small $D$ limit. This is shown to hold in simulations
presented in Appendix \ref{sec:Appendix:-DMFT-equations}. It breaks
down if the noise develops long-lasting correlations in time. Our
approximation will be nevertheless good for large $|N_{u}^{*}|$ and
for weak endogenous fluctuations.

We now used the relationship discussed above between $\overline{N_{u}}$
and $N_{u}^{*}$ to determine the statistics of $N_{u}^{*}$. We shall
use the term ``source'' for patches where $N_{u}^{*}>0$, and ``sink''
otherwise\footnote{The term ``source'' is used here so as to include patches (sometimes
referred to as pseudo-sinks) where a species might still receive migration
from patches with even larger $N_{u}^{\ast}$. But the contribution
of this migration is small and not required for its persistence.}. In order to understand the correlation between the sources in the
different communities, we unpack $N_{u}^{*}$ using Appendix \ref{sec:Appendix:-DMFT-equations}.
Taking the time-average is equivalent to averaging over the dynamical
noise $\xi$. Therefore, in patch $u$ for species $i$, $z_{i,u}=-\sigma S^{-1/2}\sum_{j}a_{ij,u}\,\overline{N_{j,u}}=-\sigma S^{-1/2}\sum_{j,+}a_{ij,u}\,N_{j,u}^{*}$.
The sum $\sum_{j,+}$ means that we only sum over $N_{j,u}^{*}>0$.
Here, we recall that $a_{ij,u}$ are standard random variables with
mean zero, variance one, and correlation between patches: 
\[
\mathbb{E}\left[a_{ij,u}\,a_{kl,v}\right]=\delta_{ik}\,\delta_{jl}\,\rho_{uv}
\]
where we used the Kronecker symbol $\delta_{ik}$.

Therefore: 
\begin{equation}
N_{i,u}^{*}=1-\mu\,m_{u}-\sigma S^{-1/2}\sum_{j,+}a_{ij,u}\,N_{j,u}^{*}\label{siEq:cavity}
\end{equation}
where we recall $m_{u}=\left\langle \overline{N_{i,u}}\right\rangle =\left\langle N_{i,u}^{*}\right\rangle _{+}$.
We can now compute the different moments of the multivariate Gaussian
random variable $N_{u}^{*}$, using equation (\ref{siEq:cavity}).
We obtain the closure:

\[
\begin{cases}
\mean\left[N_{u}^{*}\right]=1-\mu\,\left\langle N_{i,u}^{*}\right\rangle _{+}\\
\mathrm{covariance}\left[N_{u}^{*},N_{v}^{*}\right]=\sigma^{2}\rho_{uv}\left\langle N_{i,u}^{*}N_{i,v}^{*}\right\rangle _{+}
\end{cases}
\]

When $u=v$, as $\rho_{uu}=1$, we find the expected single community
result. In particular, $\mean\left[N_{u}^{*}\right]$ and $\mathrm{variance}\left[N_{u}^{*}\right]$
do not depend on the patch $u$.

We numerically solve the closure in a self-consistent way: start with
a guess for $\left\langle N_{i,u}^{*}\,N_{i,v}^{*}\right\rangle _{+}$,
and then (1) Produce many samples of the vector $N_{u=1..M}^{*}$
and (2) calculate the next estimate for $\left\langle N_{i,u}^{*}\,N_{i,v}^{*}\right\rangle _{+}$,
by averaging only over $N_{i,u}^{*}$ and $N_{i,v}^{*}$ that are
both positive. For stability of this numerical scheme, we only replace
half the samples at each iteration. We use $10^{5}$ samples and 1000
iterations. The algorithm is always found to converge to the same
solution.

Given $\mathrm{covariance}\left[N_{u}^{*},N_{v}^{*}\right]$, the
distribution of $N_{i,u}^{*}$ is completely specified: it is a multivariate
Gaussian in $u$, has the single-patch statistics of a single community,
and a known covariance between patches. The solution can then also
give the distribution of the number of sourcing patches.

\begin{figure}
\begin{centering}
\includegraphics[width=1\columnwidth]{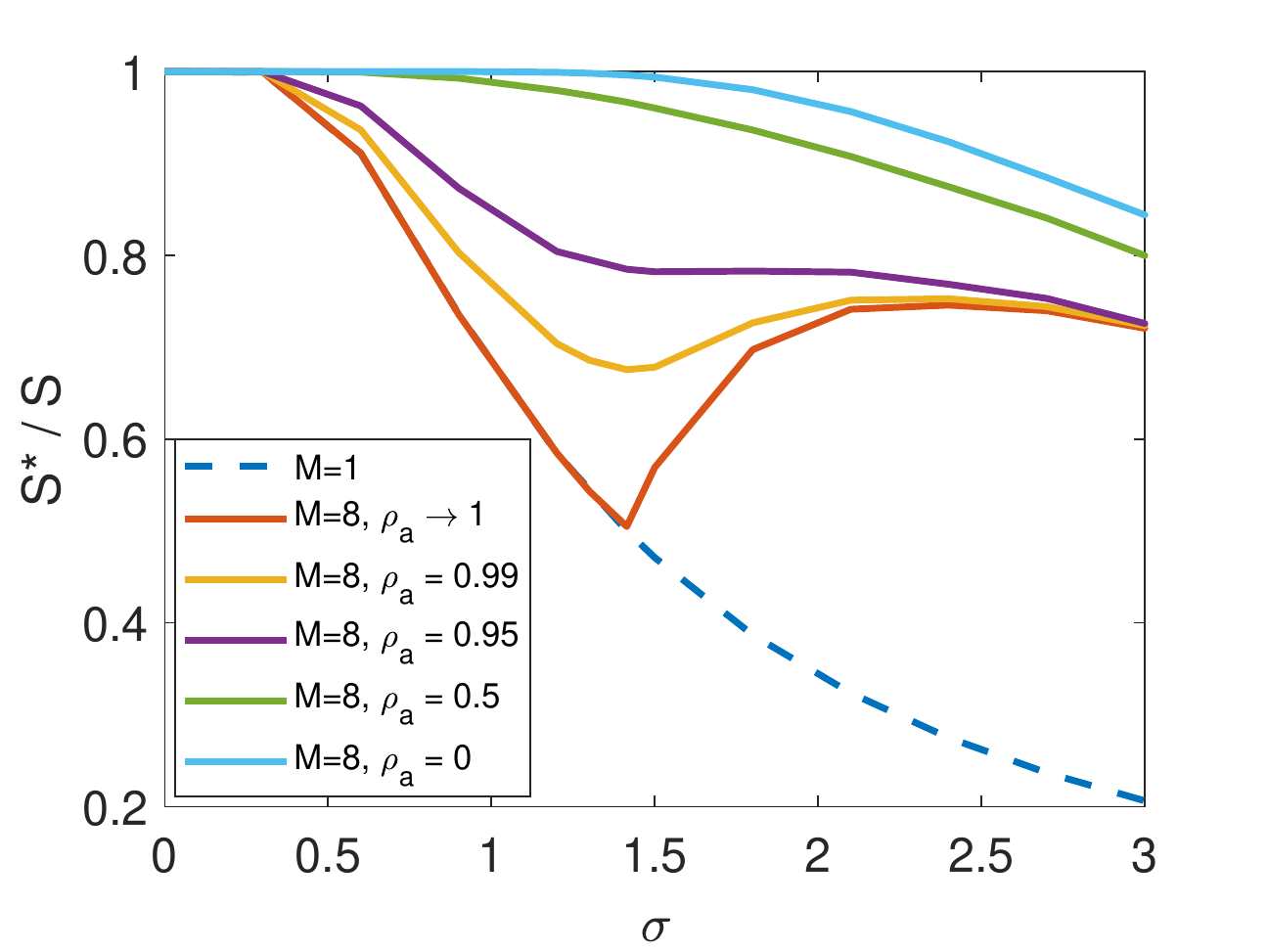} 
\par\end{centering}
\caption{Theoretical predictions for the diversity as a function of $\sigma$
for $M=1,8$ patches, $\rho_{a}=0,0.5,0.95$ and $\rho_{a}\rightarrow1$.
\label{fig:diversity_predictions}}
\end{figure}

In addition, we can compute the correlation coefficient $\rho_{N^{*}}$
of the $N_{u}^{*}$'s. We use here our simple case of a uniform correlation
$\rho_{a}$ between patches $\rho_{uv}=\rho_{a}+(1-\rho_{a})\delta_{uv}$.
We introduce the notation $\rho_{a}$ instead of `$\rho$' in this
section in order to avoid confusion with $\rho_{N^{*}}$.
\[
\rho_{N^{*}}\equiv\frac{\mathrm{covariance}\left[N_{u}^{*},N_{v}^{*}\right]}{\mathrm{variance}\left[N_{u}^{*}\right]}=\rho_{a}\frac{\left\langle N_{i,u}^{*}\,N_{i,v}^{*}\right\rangle _{+}}{\left\langle {N_{i,u}^{*}}^{2}\right\rangle _{+}}\ 
\]
The results are surprising: even when $\rho_{a}\rightarrow1$, the
overlap between communities is not perfect ($\rho_{N^{*}}<1$), so
the total diversity is larger than the one in each patch. This happens
\emph{exactly at} the transition to chaos at $\sigma_{c}=\sqrt{2}$,
see Fig. \ref{fig:diversity_predictions}.

\begin{figure}
\begin{centering}
\includegraphics[width=1\columnwidth]{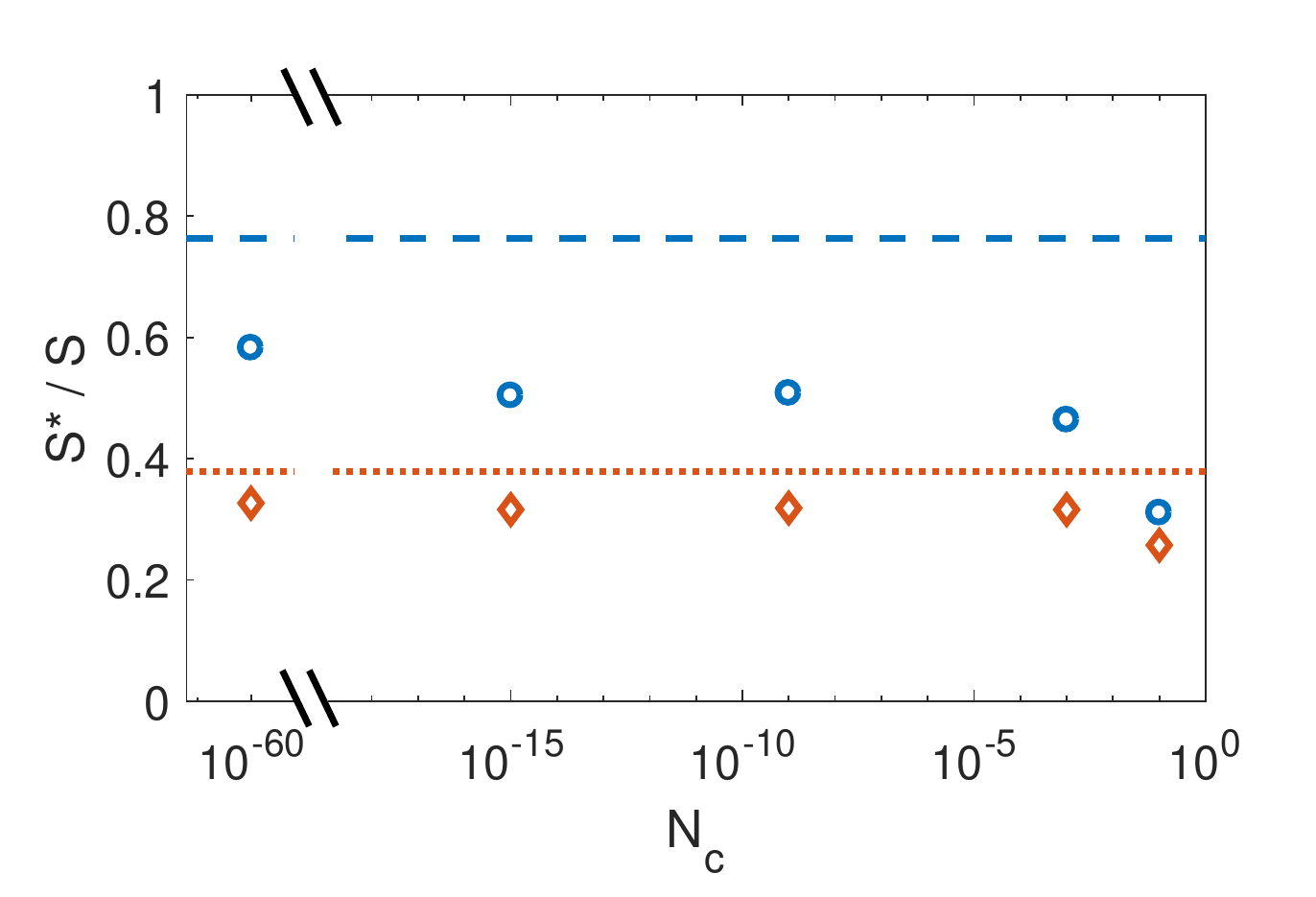}
\par\end{centering}
\caption{The fraction of persistent species $S^{*}/S$ (circles) is compared
to theoretical bound (blue dashed line), for different values of $N_{c}$.
Also shown is the fraction of species above $N_{\mathrm{eff}}^{*}>0.2$,
compared to the theoretical bound for that (red dotted line), showing
better agreement than for the full diversity. Simulations use the
same parameters as in Fig. \ref{fig:Distribution-of-Neff}, but with
a range of values for $N_{c}$ (Fig. \ref{fig:Distribution-of-Neff}
corresponds to the points at $N_{c}=10^{-15}$).\label{fig:diversity_large_Neff}}
\end{figure}

\begin{figure*}
\begin{centering}
\includegraphics[width=1.6\columnwidth]{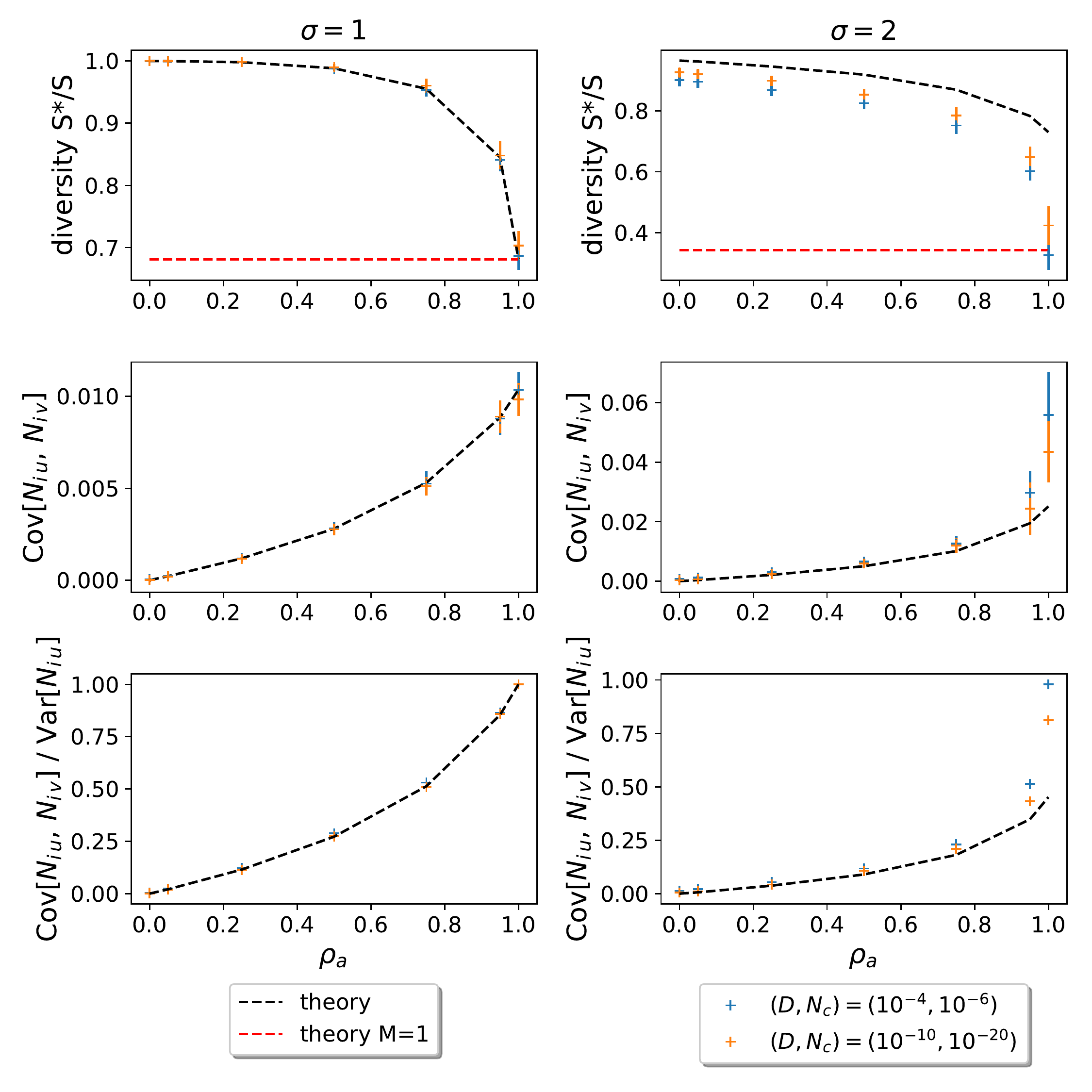}
\par\end{centering}
\caption{Numerical checks of the theoretical predictions. From top to bottom,
we consider three different observables: the diversity, the covariance
in the abundances across distinct patches, and this covariance rescaled
by the one patch variance. By varying $\sigma$, we can control the
state of the system: on the left ($\sigma=1$), we show the results
for fixed points; on the right ($\sigma=2>\sqrt{2}$), we show the
results for persistent dynamical fluctuations. In dotted lines, we
plot the theory predictions, as functions of the correlation between
patches' interactions $\rho_{a}$. We compare them to simulations
with parameters $(S,M,\mu)=(400,8,10)$, and obtained by simulations
run until final time $t_{f}=10^{4}$. We eventually vary the couple
$(D,N_{c})$. We use 50 distinct samples of the simulations for each
combination of parameters, in order to get error bars and relevant
statistics. The cut-off is implemented via patch-wise extinctions
when the abundance goes below the threshold in each particular patch,
in which case migration out of the patch is turned off while still
allowing inward migrations.\protect \protect \\
On the left side, we can see that the theory is exact in the fixed
point regime. In this regime, as $\rho_{a}\to1$, the predictions
are equivalent to the one patch $M=1$ theory, as all patches are
the same. In the persistent fluctuation state, the theory is only
a good approximation. More precisely, the predictions become more
accurate as $D$ and $N_{c}$ go to zero, as expected. In addition,
the agreement gets worse when $\rho_{a}\to1$, because synchronization
can occur. \protect \protect \\
 In the top right figure, we show that the prediction for diversity
is an upper bound. In the bottom right figure, we see that indeed
the prediction for $\rho_{n}$ is still far from $1$ when $\rho_{a}\to1$,
for the values of $D,N_{c}$ used in the simulations.\label{fig:numerical_checks}}
\end{figure*}

On Fig. \ref{fig:numerical_checks}, we compare the theory predictions
to simulations. In terms of diversity, the theory appears to give
an upper bound to the simulations. The difference becomes larger at
higher values of $\sigma$, and for $\rho_{a}$ closer to one. To
look further into this difference, it is useful to study diversity
as a function of the value of $N_{\mathrm{eff}}^{*}$ of each species.
As shown in Fig. \ref{fig:Distribution-of-Neff} in the main text,
most of the difference in diversity is due to low values of $N_{\mathrm{eff}}^{*}$,
which are precisely the species that are more likely to go extinct,
with good agreement with theory at higher values of $N_{\mathrm{eff}}^{*}$.
This is demonstrated in Fig. \ref{fig:diversity_large_Neff}, which
shows that the theoretical prediction for the number of species with
$N_{\mathrm{eff}}^{*}>0.2$ is closer to simulation results than the
predictions for total diversity. At the moment we do not know if remaining
differences are because the theoretical argument is only approximate,
or whether in principle, with exceedingly low values of $N_{c}$ and
$D$, it could be approached by simulations for any $\sigma$.

To find the boundary of parameter space where fixed points loose their
stability and the system becomes chaotic, we look at the linear stability
of persistent species. When $D$ is small, the species that are not
sourced in each patch do not affect the stability, and so the question
simplifies to single patch stability, which when $\corr\left[A_{ij},A_{ji}\right]=0$,
results in $\sigma_{c}=\sqrt{2}$ and with $1/2$ of the species being
sourced in each patch \cite{bunin_ecological_2017}.\\

\section{Single patch ($M=1$)\label{sec:Appendix:-single-patch}}

\begin{figure}
\centering{}\includegraphics[width=1\columnwidth]{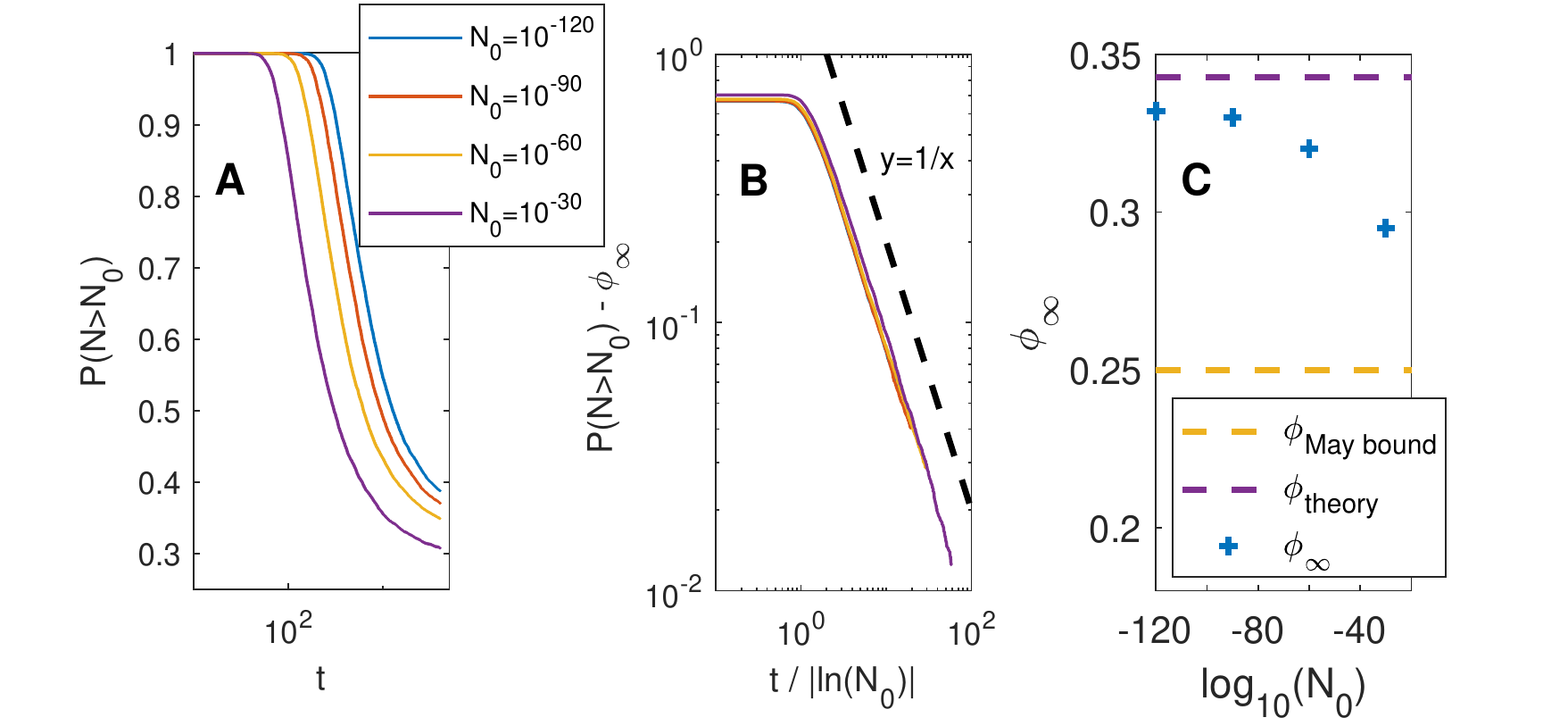}\caption{DMFT numerics for a single patch, $M=1$, showing that chaos is in
principle possible here, although for unrealistic values of model
parameters. (A) The fraction of species above different values of
$N_{0}$, $P\left(N>N_{0}\right)$ is plotted as a function of time,
for different values of $N_{0}$. (B) The curves for different $N_{0}$
collapse when $P\left(N>N_{0}\right)-\phi_{\infty}\left(N_{0}\right)\sim\left|\ln N_{0}\right|/t$.
Here $\phi_{\infty}\left(N_{0}\right)$ is a fitted parameter, the
extrapolated value of $P\left(N>N_{0}\right)$ at long times. (C)
The values of $\phi_{\infty}\left(N_{0}\right)$ are well above the
linear stability bound (``May bound''), and at (very) low $N_{0}$
come quite close to the theoretical maximal value for $\phi_{\infty}\left(N_{0}\right)$,
predicted in Appendix \ref{sec:Appendix:Diversity-calc}. Here $\sigma=2,\mu=10,N_{\mathrm{c}}=10^{-120}$.\label{fig:DMFT-M1}}
\end{figure}

\begin{figure}
\centering{}\includegraphics[width=1\columnwidth]{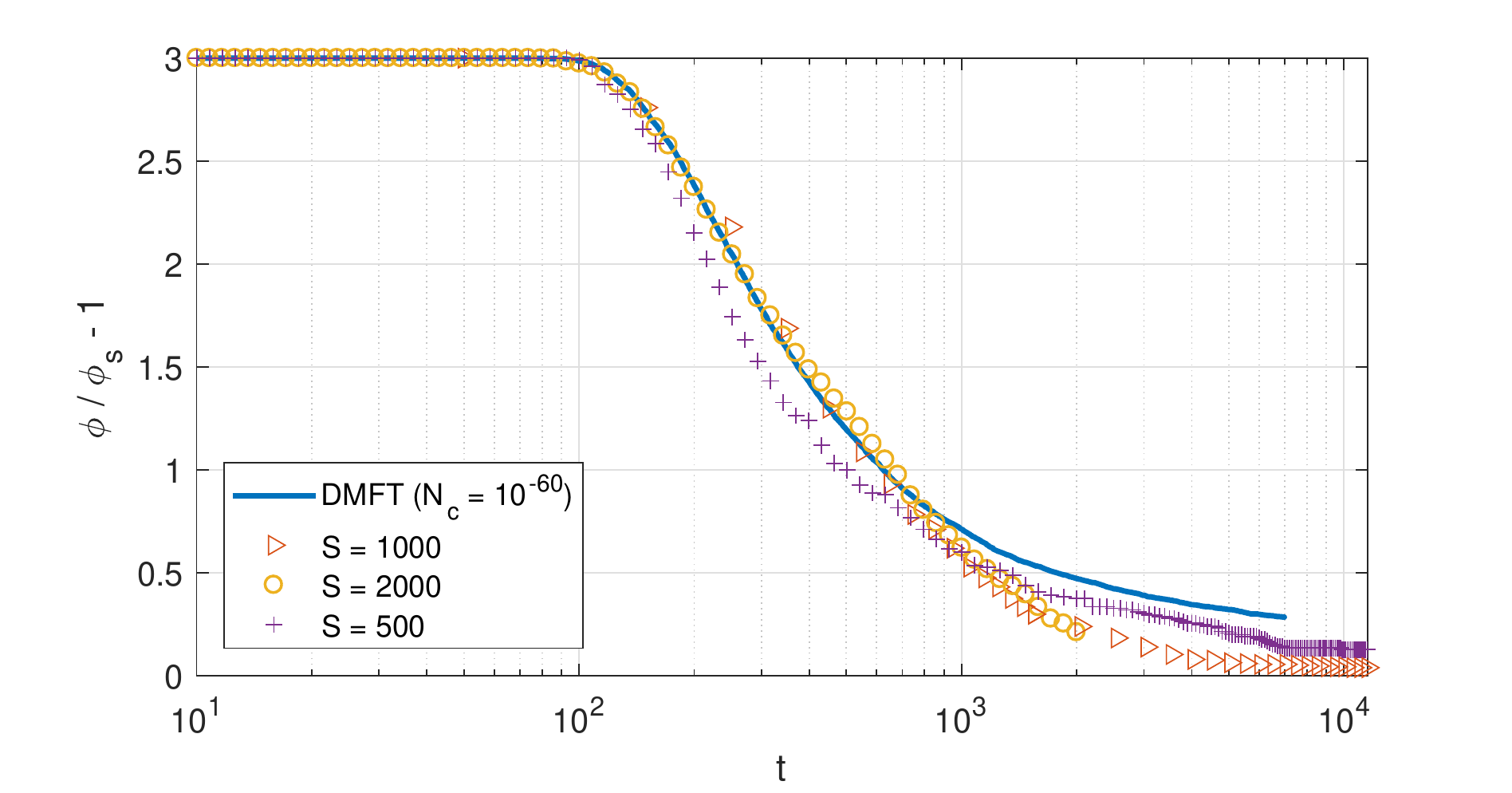}\caption{The DMFT solution and the simulations only agree up to times $t\sim10^{3}$,
after which the diversity in the simulations reduces more rapidly
and reaches a fixed point. This means that the convergence to the
DMFT solution is slow with $S$.\label{fig:DMFT-vs-simulation}}
\end{figure}

Here we show that in principle a single patch can reach and maintain
a dynamically fluctuating state. However, this requires prohibitively
large $S$, not attainable in practice. In Fig. \ref{fig:DMFT-M1}
and Fig. \ref{fig:DMFT-vs-simulation} we show results of a numerical
solution \citep{roy_numerical_2019} to the DMFT equations detailed
in Appendix \ref{sec:Appendix:-DMFT-equations}. At extremely low
values of $N_{c}$ the system appears to reach a final diversity above
the May bound and, hence, to be chaotic. DMFT however describes the
behavior in the $S\gg1$ limit. When full simulations of the model
in Eq. (\ref{eq:EOM}) are carried out at finite $S$, they diversity
falls somewhat below the DMFT final diversity, leading to a fixed
point, rather than a chaotic state, see Fig. \ref{fig:DMFT-vs-simulation}.
This finite-size correction to the DMFT result are important since
they show that maintaining a dynamically fluctuating state for realistic
values of $S$ is not possible for $M=1$.

\section{Correlations of interactions in a pair of species\label{sec:Appendix:gamma_neq_0}}

\begin{figure}
\centering{}\includegraphics[width=1\columnwidth]{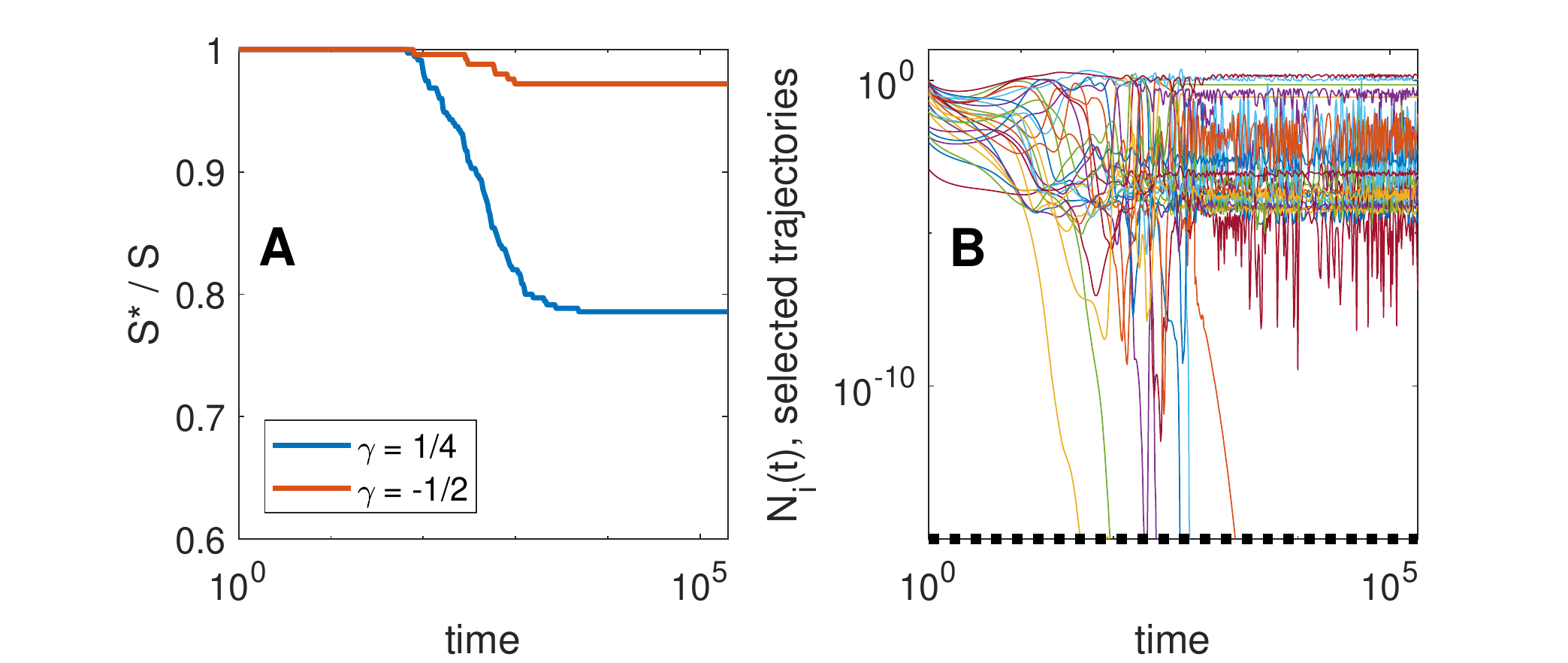}\caption{(A) The diversity $S^{*}\left(t\right)/S$ for two runs with $\gamma\equiv\protect\corr\left[A_{ij,u},A_{ji,u}\right]\protect\ne0$.
(B) Selected trajectories of $N_{i,u}\left(t\right)$ for the run
with $\gamma=1/4$.\label{fig:gam_neq_0}}
\end{figure}

In the main text we assumed that $A_{ij,u}$ is sampled independently
from $A_{ji,u}$. Here we show that the long-lived endogenous fluctuations
can be found even if this assumption is relaxed. For this purpose,
we consider a symmetric network of non-zero $A_{ij,u}$, namely $A_{ij,u}\ne0$
if and only if $A_{ji,u}$. We define $\gamma$ the correlation of
the non-zero elements $\gamma=\corr\left[A_{ij,u},\,A_{ji,u}\right]_{A_{ij,u}\ne0}$.
Fig. \ref{fig:gam_neq_0} shows two simulations, one with $\gamma>0$
and the other with $\gamma<0$. In both cases the system relaxes to
a long-lived state with fluctuating abundances, without further loss
of diversity up to time $2\cdot10^{5}$. They are intended solely
to demonstrate that conditions with $\gamma\neq0$ exist, rather than
a systematic exploration of such cases.

The parameters for the simulations (using the notation of Appendix
\ref{sec:Appendix:-model-parameters}) are the following:

Run with positive $\gamma$: $\gamma=1/4$, $S=350$, $\mean\left(A_{ij,u}\right)=0.075$,
$\std\left(A_{ij,u}\right)=0.175$, $c=0.357$, $M=8$, $d=10^{-3}$,
$\rho=0$, $N_{c}=10^{-15}$.

Run with negative $\gamma$: $\gamma=-1/2$, $S=250$, $\mean\left(A_{ij,u}\right)=0.075$,
$\std\left(A_{ij,u}\right)=0.358$, $c=0.5$, $M=8$, $d=10^{-3}$,
$\rho=0$, $N_{c}=10^{-15}$.

\end{document}